\documentclass[5p,authoryear,10pt]{elsarticle}

\usepackage{amssymb}
\usepackage{amsmath}
\usepackage{natbib}
\usepackage[dvipsnames]{xcolor}
\usepackage{subcaption}
\usepackage{graphicx}
\usepackage{epstopdf}
\usepackage{epsfig}
\usepackage{multirow}
\usepackage{tikz}
\usepackage{nicefrac}
\usepackage{arydshln}
\usepackage{url}
\usepackage{blkarray}
\usepackage{hyphenat} 
\usepackage{lipsum}
\usepackage{caption}
\usepackage{booktabs}
\usepackage{soul}
\usepackage{multirow}
\usepackage{lineno}
\usetikzlibrary{arrows,calc,intersections,positioning,arrows,decorations.pathmorphing,decorations.markings,shapes}%

\newcommand{\RN}[1]{%
  \textup{\uppercase\expandafter{\romannumeral#1}}%
}
\journal{Icarus}

\begin{document}

\begin{frontmatter}

\title{Contribution of PRIDE VLBI products to the joint JUICE-Europa Clipper moons' ephemerides solution}

\author[TUDELFT]{M.S. Fayolle\corref{cor}}
\ead{m.s.fayolle-chambe@tudelft.nl}

\author[TUDELFT]{D. Dirkx}
\ead{d.dirkx@tudelft.nl}
\author[JIVE]{G. Cimo}
\ead{cimo@jive.eu}
\author[JIVE,TUDELFT]{L.I. Gurvits}
\ead{LGurvits@jive.eu}
\author[IMCCE]{V. Lainey}
\ead{Valery.Lainey@imcce.fr}
\author[TUDELFT]{P.N.A.M. Visser}
\ead{p.n.a.m.visser@tudelft.nl}

\address[TUDELFT]{Delft University of Technology, Kluyverweg 1, 2629HS Delft, The Netherlands}
\address[JIVE]{Joint Institute for VLBI ERIC, Oude Hoogeveensedijk 4, 7991PD Dwingeloo, The Netherlands}
\address[IMCCE]{IMCCE, Observatoire de Paris, 77 Av. Denfert-Rochereau, 75014, Paris, France}

\begin{abstract}
In the coming decade, JUICE and Europa Clipper radio-science will yield the most accurate estimation to date of the Galilean moons' physical parameters and ephemerides. JUICE's PRIDE (Planetary Radio Interferometry and Doppler Experiment) will help achieve such a solution by providing VLBI (Very Long Baseline Interferometry) observations of the spacecraft's lateral position, complementing nominal radio-science measurements.
In this paper, we quantify how PRIDE VLBI can contribute to the moons' ephemerides determination, in terms of attainable solution improvement and validation opportunities. 
To this end, we simulated VLBI data for JUICE, but also investigated the possibility to perform simultaneous tracking of JUICE and Europa Clipper, thus ultimately generating both single- and dual-spacecraft VLBI. We considered various tracking and data quality scenarios for both VLBI types, and compared the formal uncertainties provided by covariance analyses with and without VLBI. These analyses were performed for both global and local (i.e. per-flyby) estimations of the moons' states, as eventually achieving a global solution first requires proceeding arc-per-arc. 
We showed that both single- and multi-spacecraft VLBI measurements only bring limited improvement to the global state estimation, but significantly contribute to the moons' normal points (i.e. local states at flyby times), most notably in the out-of-plane direction. 
Additionally, we designed a validation plan exploiting PRIDE VLBI to progressively validate the classical radio-science solution, whose robustness and statistical realism is sensitive to modelling inconsistencies. By improving the local state estimations and offering various validation opportunities, PRIDE will be invaluable in overcoming possible dynamical challenges. It can therefore play a key role in reconstructing a global solution for the Galilean moons' dynamics with the uncertainty levels promised by JUICE-Europa Clipper analyses. This, in turn, is critical to the accurate characterisation of tidal dissipation in the Jovian system, holding the key to the long-term evolution of the Galilean moons. 

\end{abstract}


\end{frontmatter}


\definecolor{color1}{RGB}{2, 81, 158}
\definecolor{color2}{RGB}{42, 193, 176}
\definecolor{color3}{RGB}{153, 61, 90}
\definecolor{color4}{RGB}{205, 119, 66}

\section{Introduction} \label{sec:introduction}

In the 2030s, both ESA's JUpiter ICY moons Explorer (JUICE) and NASA's Europa Clipper spacecraft will study Jupiter's Galilean satellites \citep{grasset2013, witasse2024, pappalardo2021}. They will perform a series of flybys around these moons, with a strong focus on Callisto (JUICE) and Europa (Europa Clipper), followed by an at least 9-month orbital phase around Ganymede for JUICE. The strong interest in the Galilean system was strengthened from the Galileo mission, with the detection, either tentative or confirmed, of subsurface oceans of liquid water below the icy crust of the three outermost satellites (Europa, Ganymede, and Callisto) \citep{khurana1998,kivelson2000,kivelson2002}. Both JUICE and Europa Clipper missions are specifically designed to confirm the findings of the Galileo mission, and provide the most detailed characterisation to date of the moons' hydrospheres \citep{petricca2023,roberts2023}.

As part of their scientific objectives, data from JUICE and Europa Clipper will further constrain the formation and long-term evolution of the Galilean system, a critical step to understand how the moons' internal oceans could have formed and survived until present-day. Our understanding of the system's thermal-orbital evolution indeed remains incomplete, with fundamental questions still open regarding the history of the Laplace resonance \citep{yoder1979,greenberg1987} and the possibility of a rapid migration of Callisto's orbit if caught in a resonance-locking mechanism \citep{lari2023}. Answering those will require a better understanding of tidal dissipation mechanisms, which govern the moons' orbital migration \citep[e.g.][]{lainey2009,lainey2020} and heats up their interiors \citep{nimmo2016}. 
The Galilean system can moreover be seen as a miniature version of the Solar System. Understanding its formation and history will therefore bring invaluable insights into planetary systems evolution in general \citep[e.g.][]{deienno2014,heller2015}.  

The moons' current orbits result from these long-term evolution processes, and therefore bear witness of the satellites' orbital and interior history. Improving our ephemerides solutions for the Galilean satellites is thus a natural way to gain insights into the system's thermal-orbital evolution. In planetary space missions such as JUICE and Europa Clipper, this is primarily achieved by extracting the dynamical signatures of the Galilean satellites from the radiometric tracking measurements of the spacecraft during their close encounters with the moons (flybys or orbital phase of Ganymede). 

For this purpose, JUICE will benefit from a dedicated radio-science instrument, 3GM (Gravity \& Geophysics of Jupiter and Galilean Moons, \citealt{iess2023}), supplemented by the High Accuracy Accelerometer (HAA) which will eliminate the effects of non-conservative perturbations (primarily propellant sloshing) in the radio-science measurements. Europa Clipper, on the other hand, will rely on the spacecraft's nominal tracking and communication radio capabilities \citep{mazarico2023}. The potential of classical radio-science observables from both JUICE and Europa Clipper has already been demonstrated for ephemerides determination applications \citep{magnanini2021,magnanini2023,fayolle2022}. These measurements are nonetheless limited by the observation geometry: range and Doppler data mostly constrain the spacecraft's motion (position and velocity, respectively) in the line-of-sight direction. 

To alleviate this limitation, JUICE will take advantage of an additional support experiment: the Planetary Radio Interferometry and Doppler Experiment (PRIDE), which has already been successfully used for, among others, the Huygens Probe, Venus Express, and Mars Express missions \citep{pogrebenko2004,bocanegra2018,duev2016}. PRIDE provides phase-referenced VLBI (Very Long Baseline Interferometry) measurements of the spacecraft's position. This is achieved by simultaneously detecting the signal transmitted by JUICE with several ground-based radio telescopes, while nodding between the target (i.e. spacecraft) and a nearby stable background radio source used as a phase calibrator. This allows PRIDE to accurately reconstruct the lateral position of the spacecraft in the ICRF (International Celestial Reference Frame), providing the missing information on JUICE's position components orthogonal to the line-of-sight. Because of the strong geometrical complementarity with the range and Doppler measurements, PRIDE VLBI data are expected to significantly help constraining the Galilean moons' dynamics. A more detailed discussion on the main differences and advantages of the PRIDE phase-referencing technique with respect to classical Delta-DOR VLBI can be found in \cite{gurvits2023}.

The contribution of JUICE-PRIDE to the ephemerides solution has already been investigated by \cite{dirkx2017}. As expected, VLBI measurements were found to mostly improve the estimation of the out-of-plane position of Jupiter and its moons. However, this previous analysis, performed with an earlier version of the JUICE trajectory, focussed on a JUICE-only radio-science solution. In addition, the methodology underlying this study, while sufficient for a preliminary analysis, did not properly capture the dynamical interactions between the spacecraft, moons and Jupiter in its uncertainty quantification. Since this first study, the expected simultaneous presence of the JUICE and Europa Clipper spacecraft in the Jovian system has radically changed the picture, the synergy between their Jovian tours greatly benefiting the ephemerides estimation. Thanks to this unique dual-mission configuration, joint JUICE-Europa Clipper analyses indeed achieve significantly more accurate and stable solutions than previous single-mission studies \citep{magnanini2023}. 

Investigating the potential of PRIDE as a powerful validation experiment, and as an additional data set to obtain a robust and stable solution, then becomes critical. The extremely low uncertainty levels predicted to be achievable by existing simulations \citep[e.g.][]{fayolle2022,magnanini2023} will indeed be extremely difficult to achieve in practice. Previous attempts to reconstruct a consistent, global solution for the motion of natural satellites from a series of radio-science flybys, in the context of the Cassini mission, have proven extremely sensitive to dynamical modelling issues, sometimes preventing or complicating the obtention of a reliable coupled solution \citep{durante2019,zannoni2020,jacobson2022}. Such issues not only impede our capability to attain a global solution, but also our assessment of the statistical realism of the obtained uncertainties (if a solution is nonetheless achieved), therefore obscuring their interpretation.

The extremely accurate radio-science data from JUICE and Europa Clipper will impose an even more stringent requirement on the consistency of our dynamical models. Similar issues as for the Cassini case are therefore expected to arise \citep[e.g.][]{dirkx2017,fayolle2022}. Overcoming these modelling challenges requires proceeding gradually, by first performing local state estimations to gradually reconstruct a coherent, global solution for the moons' ephemerides. Such a progressive approach starting from local orbit determinations is anyway typical for radiometric tracking-based analyses, where we first need to obtain an accurate spacecraft orbit solution for each flyby. By providing completely independent measurements of the spacecraft position, PRIDE will be instrumental to this step-by-step reconstruction of a robust global solution for both the spacecraft and moons' dynamics. This additional data set will be extremely valuable to validate the solutions based on range and Doppler data, assess the realism of their uncertainties, as well as to detect and identify potential modelling issues.
 
Furthermore, the presence of two in-system spacecraft opens novel, unique opportunities to perform simultaneous VLBI tracking of both JUICE and Europa Clipper. This tracking configuration, referred to as multi-spacecraft VLBI in the following, will provide extremely accurate measurements of the relative angular position between the two spacecraft. These can translate into constraints on the relative position of the Galilean moons with respect to one another, as most of JUICE's flybys occur around Ganymede and Callisto, while Europa Clipper focusses on Europa. These unique observations therefore have the potential to greatly help constraining the strongly coupled dynamics of the Galilean system.

In light of the above, this paper analyses the contribution of various PRIDE VLBI products to the moons' ephemerides determination from JUICE and Europa Clipper radio-science. We specifically quantify how much VLBI measurements can improve the solution obtained from Doppler and range data, both for local and global estimations of the moons' orbits. To this end, we pay particular attention to the error budgets of our VLBI observables, using more detailed and realistic random and systematic noises than in \citep{dirkx2017}. We moreover identify promising opportunities to perform multi-spacecraft tracking and assess the contribution of the resulting observables to the ephemerides solution. Finally, several validation strategies enabled by the PRIDE VLBI technique are explored. We discuss their potential, and investigate their upcoming role in the progressive reconstruction of a statistically consistent solution for the Galilean moons' ephemerides from JUICE and Europa Clipper data.

We first describe our simulated VLBI observables in Section \ref{sec:vlbi}, before presenting the details of our joint JUICE-Europa Clipper estimation setup in Section \ref{sec:estimation}. The underlying numerical model used for the estimation is extended from \cite{fayolle2022}. Sections \ref{sec:resultsVlbi} and \ref{sec:resultsMsVlbi} then present the results obtained when adding single- and multi-spacecraft VLBI measurements, respectively, to the joint JUICE-Europa Clipper ephemerides solution. Finally, Section \ref{sec:validation} discusses the various validation opportunities offered by the PRIDE VLBI technique and Section \ref{sec:conclusion} provides the main conclusions of our analyses.

\section{VLBI observables} \label{sec:vlbi}

Our analyses will rely on simulated VLBI observables to quantify the expected PRIDE contribution to the ephemerides solution for the Galilean moons. In this perspective, this section presents our simulated VLBI measurements, starting with describing the adopted error budget and the search process for the VLBI phase calibrators in Sections \ref{sec:vlbiErrorBudget} and \ref{sec:calibrators}, respectively. We then discuss the conditions and opportunities to perform multi-spacecraft VLBI tracking between the JUICE and Europa Clipper spacecraft in Section \ref{sec:msVlbi}. 

\subsection{Error budget for phase-referencing VLBI} \label{sec:vlbiErrorBudget}

\begin{table*}[tbp!]
	\caption{Error budget from past VLBI measurements, and selected noise levels for our simulations. Some numbers are struck out to indicate that they are deemed not representative of the typical VLBI accuracy.}
	\label{tab:errorBudget}
	\centering
    \resizebox{\textwidth}{!}{
	\begin{tabular}{l c c l}
		\textbf{Data} & $\boldsymbol{1\sigma(\alpha)}$ \textbf{[mas]}& $\boldsymbol{1\sigma(\delta)}$ \textbf{[mas]} & \textbf{Comment} \\
		\hline
		Pre-fit residuals for VEX & 0.09 & \st{0.25} & Unfavourable target's declination \\
		\cite{duev2012} &  &  &  (between -11 and -13 deg)\\
		\hline
		Pre-fit residuals for MEX & 0.03 & 0.06 & \\ 
		\cite{duev2016} & & & \\ 
		\hline
		Post-fit residuals for Cassini & 0.12 & 0.18 & After removing calibrator's and spacecraft's \\
		\cite{jones2020} & & & position uncertainty ($\sim$50\% of the residuals' rms) \\ 
		\hline
		Simulation-based errors & [0.03 ; 0.045] & [0.05 ; 0.13] & Total tropospheric effect \\
		\cite{pradel2006} & & & \\
		\hline
		& & & \\
		\textbf{Selected noise} & \multicolumn{2}{c}{$\boldsymbol{1\sigma(\alpha)}$} & \multicolumn{1}{c}{$\boldsymbol{1\sigma(\delta)}$}  \\
		\hline 
		Poor VLBI case & \multicolumn{2}{c}{0.12 mas $\approx$ 0.6 nrad} & \multicolumn{1}{c}{0.18 mas $\approx$ 0.9 nrad} \\
		Good VLBI case & \multicolumn{2}{c}{0.04 mas $\approx$ 0.2 nrad} & \multicolumn{1}{c}{0.06 mas $\approx$ 0.3 nrad} \\
		Ka-band case & \multicolumn{2}{c}{0.02 mas $\approx$ 0.1 nrad} & \multicolumn{1}{c}{0.03 mas $\approx$ 0.15 nrad} \\
		\hline
	\end{tabular}}
\end{table*}

To make our simulations as realistic as possible, the noise budget assigned to VLBI simulated data was designed based on past measurements. The main error sources are media propagation delays (interplanetary plasma, troposphere, and ionosphere), instrumental signal delays, clock offsets and instabilities, signal-to-noise ratio (SNR) of the spacecraft's signal and calibrator's broadband emission, as well as uncertainties in ground stations' coordinates, and Earth's orientation parameters \citep[][]{pradel2006}. Moreover, the quality of past VLBI data is often assessed by analysing post-fit residuals, which are however not only sensitive to the accuracy of the VLBI measurements but also affected by the quality of the orbit determination solution and by the position uncertainty of the calibrator in the ICRF.  

Phase-referencing VLBI was conducted with both the Venus Express (VEX) and Mars Express (MEX) spacecraft as observing targets. For the former, the analysis of pre-fit residuals between the VLBI data points and the a priori trajectory of the spacecraft revealed a large discrepancy between right ascension and declination \citep{duev2012}. The low declination of the MEX spacecraft (ranging from $-11$ deg to $-13$ deg), combined with a relatively large separation (2.5 deg) with respect to the phase calibrator, resulted in the poor cancellation of tropospheric and ionospheric effects, mostly translating in a large declination error. 
The MEX VLBI measurements, on the other hand, show smaller pre-fit residuals: the median values of the rms residuals are 0.03 mas and 0.06 mas\footnote{1 mas  = 4.84 nrad} in right ascension and declination, respectively, with a 2-min integration time \citep{duev2016}. 

Furthermore, \cite{jones2020} provide an overview of the VLBI measurements of the Cassini spacecraft over the entire mission duration (2004-2017). After removing outliers due to poor a priori orbit determination solution for Cassini and/or large separation between the spacecraft and calibrator (larger than 7 deg), the rms residuals are 0.24 mas and 0.36 mas in right ascension and declination, respectively. The orbit determination error and the uncertainty in the calibrators' ICRF positions can however account for half of these residuals. Both error sources are not inherently related to the VLBI measurement accuracy, and they will be accounted for independently in our simulations. We thus consider a VLBI measurement quality of 0.6 nrad ($\sim0.12$ mas) and 0.9 nrad ($\sim0.18$ mas) for Cassini's VLBI data in right ascension and declination, respectively.   

VLBI astrometry of the Juno spacecraft during the early phase of the mission has also been published, yielding rms (post-fit) residuals of 0.4 mas and 0.6 mas in right ascension and declination, respectively \citep{jones2019,park2021}. These residuals are larger than for Cassini, due to the poorer quality of Juno's a priori orbit solution available at the beginning of the mission, as well as large calibrator position errors for some epochs \citep{jones2019}. The few published Juno VLBI measurements were thus not considered representative of the accuracy typically expected from VLBI phase-referencing tracking.

Based on these existing measurements, and excluding the pessimistic errors in declination obtained with VEX, we selected two different Gaussian random noise budgets for our simulated VLBI observables (see Table \ref{tab:errorBudget}):
\begin{itemize}
	\item Poor VLBI noise case: $\sigma(\alpha)$ = 0.12 mas and $\sigma(\delta)$ = 0.18 mas, based on Cassini VLBI post-fit residuals after removing the estimated contribution of the calibrators' positions ;
	\item Good VLBI noise case: $\sigma(\alpha)$ = 0.04 mas\footnote{The error in right ascension was set to 0.04 mas instead of 0.03 mas to keep the same ratio between the good and poor VLBI noises in both right ascension and declination.} and $\sigma(\delta)$ = 0.06 mas, consistent with MEX VLBI measurements and with the minimum tropospheric effect errors.
\end{itemize}
The above only encompasses random error sources, and does not account for the systematic bias induced by an error in the calibrator's ICRF position, which will be addressed in Section \ref{sec:calibrators}.

These error levels are consistent with the simulation-based analysis of VLBI systematic errors in \cite{pradel2006}. They indeed identified wet tropospheric effects as the dominant error source, apart from the uncertainty in the calibrator's position, which in our case is treated as a separate bias (see Section \ref{sec:calibrators}). The error due to the total tropospheric effect was found comprised between 0.03 and 0.045 mas in right ascension, and between 0.05 and 0.13 mas in declination. These values give an indication of the minimum noise level that can be expected for VLBI measurements, and are in line with our good VLBI case. Moreover, the existing VLBI measurements on which we based our error budget did not benefit from dual-frequency calibration techniques to cancel ionospheric effects, nor from water vapor radiometers for wet tropospheric delay calibration. The quality of these data points can thus be considered rather conservative with respect to the highest accuracy achievable with the phase-referencing VLBI technique \citep{jones2020}.

It should also be noted that the random errors in VLBI observables cannot be perfectly represented by purely uncorrelated white noise. In practice, some uncertainty sources (e.g. atmospheric delays) are time-dependent, limiting how frequently \textit{independent} (i.e. uncorrelated) VLBI data points can be obtained. In the following, we will therefore consider different VLBI measurement cadences to account for this and avoid overestimating the data volume and information content of the VLBI data set (see Section \ref{sec:simulatedTracking}).

The two VLBI noise budgets mentioned above rely on past X-band VLBI measurements and thus indirectly assume tracking at similar frequencies. However, JUICE is also equipped with Ka-band tracking capabilities. While no existing VLBI data at such frequencies can be exploited to derive realistic noise budgets, a factor two to four improvement can be theoretically expected between X- and Ka-band measurements. We thus considered an additional case, referred to as Ka-band case (see Table \ref{tab:errorBudget}), with VLBI noise level set to half their X-band values in the best case scenario. The actual feasibility of Ka-band VLBI tracking will eventually depend on the availability of both suitable calibrators at these frequencies (see Section \ref{sec:calibrators}), and VLBI arrays with sufficient number of Ka-band-capable telescopes.

\subsection{Phase-referencing VLBI calibrators} \label{sec:calibrators}

\begin{figure*} [tbp!] 
	\centering
	\begin{minipage}[l]{0.49\textwidth}
		\centering
		\subcaptionbox{Uncertainties in right ascension. \protect\label{fig:biasCalibratorRa}}
		{\includegraphics[width=1.0\textwidth]{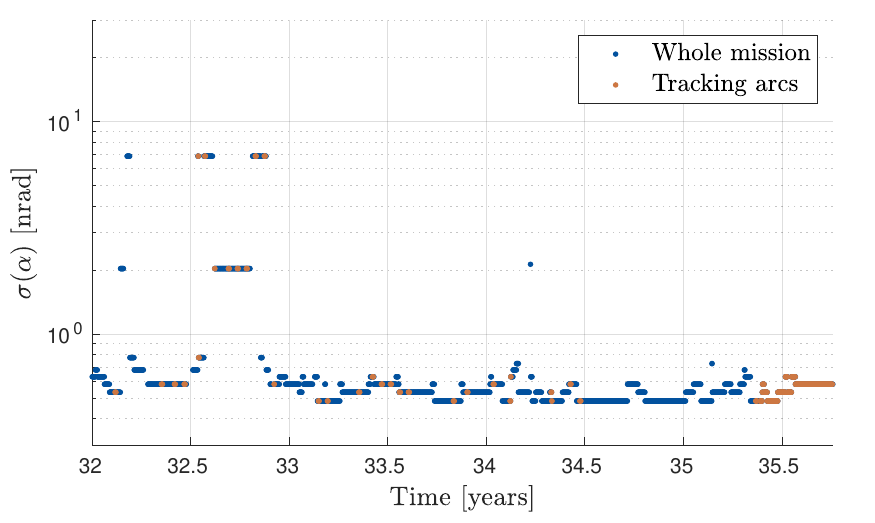}}
	\end{minipage}
	\hfill{} 
	\begin{minipage}[r]{0.49\textwidth}
		\centering
		\subcaptionbox{Uncertainties in declination. \protect\label{fig:biasCalibratorDec}}
		{\includegraphics[width=1.0\textwidth]{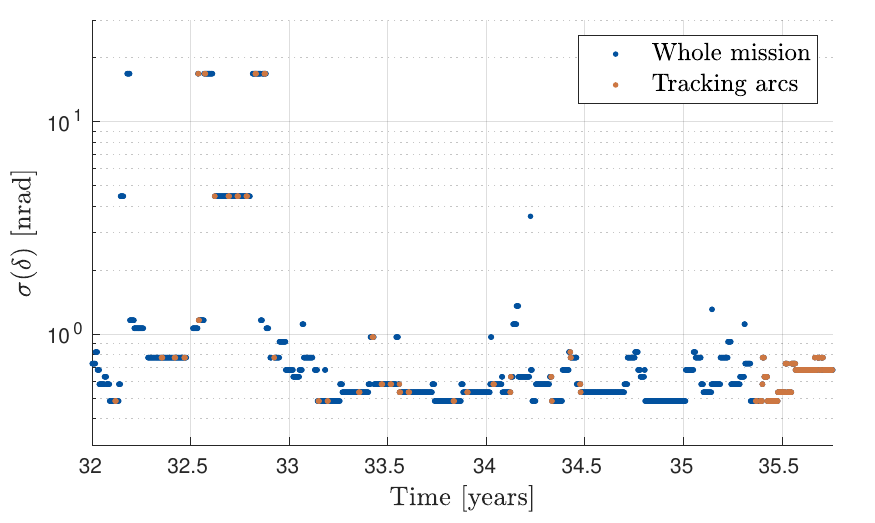}}
	\end{minipage}
	\caption{Uncertainties in the ICRF positions of the phase calibrators identified over the course of the JUICE mission, using the latest JUICE trajectory (see Section \ref{sec:dynamicalModels}). The blue dots represent all calibrators (at all epochs), while we highlighted in orange the epochs corresponding to an active tracking session during the flyby and orbital phases, when VLBI measurements are actually possible.}
	\label{fig:calibrators}
\end{figure*}

The phase-referencing VLBI technique used by PRIDE requires nodding between the target (spacecraft) and a nearby radio source, used as calibrator, to yield very accurate measurements of the target's lateral position in the ICRF. For each of our simulated VLBI data points (see Section \ref{sec:simulatedTracking}), we therefore first verified that a suitable phase calibrator is available, which implies fulfilling the following conditions. First, the radio source must have a sufficiently high total flux density at X- and Ka-bands \citep[e.g. at least $\sim$30 mJy as was the case for MEX observations in 2013,][]{duev2016}. Furthermore, the source must be compact (i.e. bright) enough for a major part of the total flux density coming from a compact, mas-scale morphology. Additionally, the angular separation between the calibrator and the spacecraft should typically be smaller than 2 deg to obtain accurate phase-calibrated measurements.

Moreover, as mentioned in Section \ref{sec:vlbiErrorBudget}, an error in the calibrator's ICRF position would introduce a systematic bias in the spacecraft's angular position derived from the VLBI observation. For each calibrator identified over JUICE's Jovian tour, we therefore extract its position uncertainty from the Radio Fundamental Catalog (rfc2023b\footnote{\url{http://astrogeo.org/rfc/}}), as shown in Figure \ref{fig:calibrators}, to be applied as a systematic bias. The influence of such biases, which vary over the mission duration as different calibrators are used, is thus directly accounted for in our estimations (see Section \ref{sec:parameters}). From Figure \ref{fig:calibrators}, the averaged uncertainty values are 0.8 nrad and 1.3 nrad in right ascension and declination, respectively. However, the calibrators' position accuracy is significantly worse between mid-2032 and 2033, due to the absence of better calibrators within 2 deg of the spacecraft. This period unfortunately overlaps with eight of JUICE's flybys, including its two flybys at Europa, and will be further discussed in our results (Section \ref{sec:resultsVlbi}).

A similar calibrator search was conducted in Ka-band, but the limited number of catalogued radio source at these frequencies yielded poor results (no suitable Ka-band calibrator during the flyby phase). Our results for the Ka-band case should therefore be treated carefully, as they depend on the hypothetical presence of a nearby appropriate calibrator. In the following, we arbitrarily used X-band calibrators for our Ka-band analyses. If the added-value of Ka-band VLBI is demonstrated, future observation campaigns to densify the Ka-band radio source background should be conducted before JUICE reaches the Jovian system.

\subsection{Multi-spacecraft in-beam measurements} \label{sec:msVlbi}

\begin{figure*}[tbp!]
	\centering
	\makebox[\textwidth][c]{\includegraphics[width=0.8\textwidth]{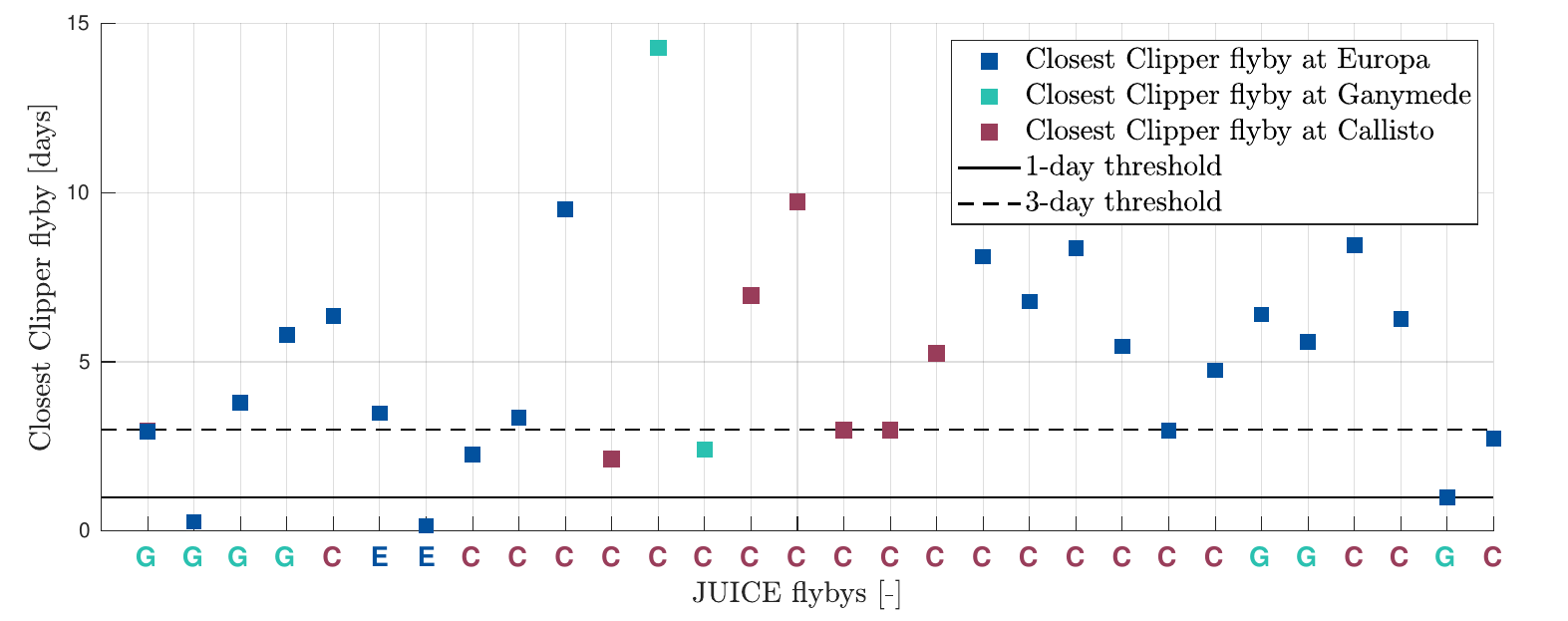}}
	\caption{Timespan between each JUICE flyby and the temporarily nearest Europa Clipper flyby, using the latest mission trajectories (see Section \ref{sec:dynamicalModels}), to identify multi-spacecraft tracking opportunities. The colours indicate around which moon each flyby is performed, and the horizontal plain and dashed lines represent a time interval limit of one and three days, respectively.}
	\label{fig:msVlbiOptions}
\end{figure*}

As mentioned in Section \ref{sec:introduction}, the simultaneous presence of JUICE and Europa Clipper in the Jovian system will make it possible to perform concurrent VLBI tracking of the two spacecraft. Such multi-spacecraft VLBI measurements have already been acquired for various Mars missions: between the Phoenix spacecraft and the Martian orbiters MRO (Mars Reconnaissance Orbiter) and Odyssey \citep{fomalont2010}, and between MEX (Mars Express), TGO (Trace Gas Orbiter), and MRO \citep{moleracalves2021}. 

Performing multi-spacecraft VLBI tracking requires the two spacecraft (here JUICE and Europa Clipper) to be concurrently transmitting, with a suitable calibrator within 2 deg of the targets, as for single-spacecraft VLBI. Moreover, for in-beam tracking to be feasible, the angular separation between the two target spacecraft should be smaller than the beam size of a typical single-dish telescope involved in the observation ($\sim$3 arcmin for a 30-m-class radio telescope observing at X-band). In addition to these feasibility requirements, some additional conditions should also be fulfilled for the multi-spacecraft VLBI measurements to significantly contribute to the moons' ephemerides estimation. The most promising opportunities indeed occur when the two spacecraft are both temporally close to an encounter (\textit{i.e} flyby) with a moon, such that the spacecraft's motions still contain signatures of the moons' dynamics. When looking for multi-spacecraft VLBI tracking opportunities, we therefore focus on combinations of two flybys, one by JUICE and one by Europa Clipper, less than three days apart. This also ensures that the (potentially relatively large) pre-encounter and clean-up manoeuvres planned three days before and after each flyby \citep{esocNav,young2019}, respectively, are excluded from the tracking arcs and do not affect the estimation. 

Based on these requirements, Figure \ref{fig:msVlbiOptions} highlights possible multi-spacecraft VLBI opportunities. In total, 11 flyby combinations meet the maximum time interval requirement of three days, as summarised in Table \ref{tab:msVlbiOptions}. Seven of them involve flybys performed around two different moons (referred to as multi-moon flyby combinations). The remaining four, on the other hand, are flybys performed at the same moon (single-moon flyby combinations), including a flyby of Europa by both spacecraft with less than four hours in-between. The potential of such tracking configurations, due to their unique geometry, to validate the radio-science solution(s) or detect dynamical modelling issues will be discussed and exploited in Section \ref{sec:validation}.

Multi-spacecraft VLBI tracking will yield very accurate measurements of the relative position of the two spacecraft in the ICRF. However, a distinction should be made between in-beam and telescope nodding phase referencing. If the two spacecraft are close enough (less than 3 arcminutes), their signals can be simultaneously tracked within the primary beam of the radio telescope. In such in-beam configuration, many systematic errors affecting the quality of the measurement cancel out \citep{majid2007,fomalont2010}. Based on previous in-beam experiments, an accuracy of 0.1 nrad could then be expected for the relative angular position measurement \citep{fomalont2010}. To be conservative, we also considered a poor accuracy case. Given the very small angular separation between the two targets and the cancellation of many measurement errors \citep{majid2007}, we used the good single-spacecraft VLBI case (see Section \ref{sec:vlbiErrorBudget}) for the poor in-beam VLBI error. 

For nodding multi-spacecraft measurements (when the two spacecraft are too far apart for in-beam tracking), the noise budget is slightly worse as the error cancellation is not as effective. For our good noise case, we used the same error levels for single-spacecraft VLBI (see Section \ref{sec:vlbiErrorBudget}). However, for the poor noise case, we set our multi-spacecraft VLBI errors halfway between single-spacecraft VLBI's best and worst cases, the latter being too pessimistic for multi-spacecraft tracking. For single-spacecraft VLBI, large errors are indeed only obtained for large angular separations, or with phase calibrators whose ICRF positions are poorly constrained, neither of these two conditions being relevant in a multi-spacecraft tracking configuration. Table \ref{tab:msVlbiErrorBudget} summarises these different noise levels for multi-spacecraft VLBI.

\begin{table}[tbp!]
	\caption{Combinations of JUICE and Europa Clipper flybys allowing for multi-spacecraft VLBI tracking.}
	\label{tab:msVlbiOptions}
	\centering
    \small
    \resizebox{0.49\textwidth}{!}{
	\begin{tabular}{c c c c c}
		\textbf{} & \textbf{JUICE} & \textbf{Clipper} & \textbf{Time} & \textbf{In-beam} \\
		\textbf{} & \textbf{flyby at} & \textbf{flyby at} & \textbf{interval [h]} & \textbf{possible} \\
		\hline
		1 & \textcolor{color2}{Ganymede} & \textcolor{color1}{Europa} & 70.7 & no \\
		2 & \textcolor{color2}{Ganymede} & \textcolor{color1}{Europa} & 6.5 & yes \\
		3 & \textcolor{color1}{Europa} & \textcolor{color1}{Europa} & 3.7 & yes \\
		4 & \textcolor{color3}{Callisto} & \textcolor{color1}{Europa} & 54.3 & yes \\
		5 & \textcolor{color3}{Callisto} & \textcolor{color3}{Callisto} & 51.3 & yes \\
		6 & \textcolor{color3}{Callisto} & \textcolor{color2}{Ganymede} & 57.8 & partially \\
		7 & \textcolor{color3}{Callisto} & \textcolor{color3}{Callisto} & 71.5 & no \\
		8 & \textcolor{color3}{Callisto} & \textcolor{color3}{Callisto} & 71.5 & no \\
		9 & \textcolor{color3}{Callisto} & \textcolor{color1}{Europa} & 71.4 & partially\\
		10 & \textcolor{color2}{Ganymede} & \textcolor{color1}{Europa} & 23.9 & no \\
		11 & \textcolor{color3}{Callisto} & \textcolor{color1}{Europa} & 65.6 & partially \\	
		\hline
	\end{tabular}}
\end{table}

\begin{table}[tbp!]
	\caption{Selected error levels for simulated multi-spacecraft VLBI measurements.}
	\label{tab:msVlbiErrorBudget}
	\centering
    \small
    \resizebox{0.49\textwidth}{!}{
	\begin{tabular}{c c c c c}
		\textbf{Selected} & \multicolumn{2}{c}{$\boldsymbol{1\sigma(\alpha)}$ \textbf{[nrad]}} & \multicolumn{2}{c}{$\boldsymbol{1\sigma(\delta)}$ \textbf{[nrad]}} \\
		\textbf{noise} & in-beam & nodding & in-beam & nodding \\
		\hline
		Poor case & 0.2 & 0.4 & 0.3 & 0.6 \\
		Good case & 0.1 & 0.2 & 0.1 & 0.3 \\
		\hline
	\end{tabular}}
\end{table}

\section{Estimation setup for joint JUICE - Europa Clipper solutions} \label{sec:estimation}

This section describes the estimation setup for our JUICE-Europa Clipper radio-science simulations, starting with the models used to propagate the dynamics of both the moons and spacecraft in Section \ref{sec:dynamicalModels}. Sections \ref{sec:simulatedTracking} and \ref{sec:estimationStrategy} then present the simulated observables and state estimation strategies applied in our analyses, respectively, before Section \ref{sec:parameters} lists the various parameters to be estimated.  

\subsection{Dynamical models} \label{sec:dynamicalModels}

In our analyses, the dynamics of all bodies involved (i.e., Jupiter, moons, spacecraft) are concurrently integrated, to ensure the complete consistency of the dynamical solutions. Following the recommendations formulated in \cite{dirkx2016} and the models used in \cite{fayolle2023}, the dynamics of the Galilean satellites were propagated in a jovicentric frame using the following set of accelerations:
\begin{itemize}
	\item mutual spherical harmonics acceleration between Jupiter and each moon, considering all zonal coefficients for Jupiter up to degree 10, and expanding the moons' gravity fields up to degree and order 2;
	\item mutual spherical harmonics acceleration between the four Galilean moons, including interactions between terms up to degree and order 2;
	\item point mass gravity from the Sun and Saturn;
	\item relativistic acceleration corrections;
	\item tidal effect on the orbit of moon $k$ due to the tides raised on Jupiter by moon $k$ (see discussion below);
	\item tidal effect on the orbit of moon $k$ due to the tides raised by Jupiter on moon $k$.	
\end{itemize}

The moons' gravity field coefficients were taken from \cite{schubert2004}, while Jupiter's gravity field was based on the current state-of-the-art model at mid-Juno mission \citep{iess2018,durante2020}. We used the latest IAU model for Jupiter's rotation \citep{archinal2018}, and the moons' rotations were assumed to be synchronous, with their long axis pointing towards the empty focus of their orbit \citep[e.g.][]{lari2018}. We considered zero obliquity for all four satellites.

We chose to directly model the effects of tides on the moons' orbits, following the formulation proposed in e.g. \cite{lari2018,lainey2019}, instead of introducing time-variation of the satellites' gravity fields due to tidal deformation. The motivation for this modelling choice is twofold. First, it circumvents the need for (near)-perfect consistency between our tidal and rotational models to accurately reproduce the effects of tides on the moons' dynamics \citep[e.g.][]{dirkx2016}. More importantly, this allows us to focus on the signature of the tidal effects present in the moons' orbits specifically, and not in the gravity field variations sensed by the spacecraft \citep[analysed in][]{magnanini2023}. This allows us to investigate how PRIDE VLBI measurements might help estimate tidal dissipation parameters via an improved determination of the moons' ephemerides.

Our estimation setup also requires propagating Jupiter's dynamics (heliocentric frame), for which the following accelerations set was considered:
\begin{itemize}
    \item mutual spherical harmonics acceleration between Jupiter and the Sun, expanding both gravity fields up to degree and order 2;
    \item point mass gravity from all planets in the Solar System and from the four Galilean satellites;
    \item relativistic acceleration corrections.
\end{itemize}
It must be noted that gravitational perturbations exerted by major belt asteroids were neglected, while they need to be accounted for in precise planetary ephemerides determination \citep{park2021,fienga2019}. However, the focus of our analysis is not on the refinement of the Jovian orbit, and we only include Jupiter in our estimations to ensure that its influence on the moons' ephemerides is considered. This makes this simplification acceptable, especially in a covariance analyses context.

Finally, the orbits of the JUICE and Europa Clipper spacecraft were propagated with respect to the central moon of each flyby and/or orbital phase, using the latest available trajectories as references\footnote{JUICE trajectory: juice\_mat\_crema\_5\_0\_20220826\_20351005\_v01 \url{https://www.cosmos.esa.int/web/spice/spice-for-juice}}\footnote{Europa Clipper trajectory: 21F31\_MEGA\_L241010\_A300411\_LP01\_V4\_postLaunch\_scpse \url{https://naif.jpl.nasa.gov/pub/naif/EUROPACLIPPER/kernels/spk/}}. 
The following set of accelerations was considered:  
\begin{itemize}
	\item spherical harmonics gravitational acceleration from Jupiter (zonal coefficients up to $J_{10}$);
	\item spherical harmonics gravitational acceleration from the central moon up to degree and order 13 (Europa), 15 (Ganymede), and 9 (Callisto) (see Section \ref{sec:parameters}); 
	\item point mass gravity from the other (non-central) Galilean moons, the Sun, and Saturn;
	\item solar radiation pressure from the Sun;
	\item arc-wise empirical accelerations, constant in the RTN (radial, tangential, normal) frame (nominal values set to zero), modelling possible accelerometer calibration errors. 
\end{itemize}

Regarding the latter, one set of empirical accelerations was considered for each flyby and for each daily arc during JUICE's GCO (Ganymede Circular Orbit)
phase. Longer arcs were however considered in-between flybys for multi-spacecraft tracking (Section \ref{sec:msVlbi}), during which daily empirical accelerations were added to modelled expected perturbations of the spacecraft's dynamics.

\subsection{Simulated radio-science observations}
\label{sec:simulatedTracking}

For our covariance analyses, we first simulated classical radio-science measurements (Doppler and range) for both JUICE and Europa Clipper. For the sake of clarity, the range and Doppler-only solution, with no VLBI included, will be referred to as the baseline solution in the rest of this paper. 

For JUICE, we assumed a X/Ka-band link and three tracking arcs of 6h each per flyby, one centered around the closest approach and the other two planned 12h before and after the flyby, following the configuration used in \cite{cappuccio2022}. We assumed that Ka-band tracking capabilities will be available for all three ESTRACK (European Space Tracking) stations involved in the tracking of the JUICE spacecraft. We therefore considered that X/Ka-link is available for all tracking arcs, in agreement with e.g., \cite{cappuccio2020,cappuccio2022,magnanini2023}. In addition, the GCO was divided in day-long arcs, with 8h of tracking per day. For each of these tracking arcs, Doppler and range data were simulated with a noise level of 12 $\mathrm{\mu}$m/s (60s of integration time) and 20 cm, respectively. Although in agreement with similar JUICE simulation analyses \citep[e.g.][]{cappuccio2022}, this range noise budget is very conservative based on BepiColombo's sub-centimeter ranging accuracy \citep{cappuccio2020b, genova2021}. 

For Europa Clipper, only Doppler measurements were simulated from the DSN (Deep Space Network) tracking stations \citep{mazarico2023}. We assumed a noise level of 0.1 mm/s during the 4h-long tracking arcs centered at each closest approach, due to the unavailability of the high gain antenna (HGA) \citep{mazarico2023}. We also considered more accurate Doppler data with a noise of 0.05 mm/s, to be acquired during the navigation passes (HGA available). These additional passes are, on average, scheduled 20h before and after each flyby \citep{magnanini2023}. 

We then also simulated single- and multi-spacecraft PRIDE VLBI observations. Since our analyses focus on the contribution of such measurements, we considered different data acquisition and noise level scenarios, varying the following settings:
\begin{itemize}
    \item VLBI random noise, using the different error budgets defined in Section \ref{sec:vlbiErrorBudget};
    \item measurement cadences (i.e. how often can an \textit{independent} VLBI data point be generated, Section \ref{sec:vlbiErrorBudget}) of 1 h, 20 min, 5 min, and 2 min;
    \item frequency of the VLBI tracking sessions during JUICE's GCO (from weekly to monthly).
\end{itemize}

\begin{figure*}[tbp!]
	\centering
	\makebox[\textwidth][c]{\includegraphics[width=0.9\textwidth]{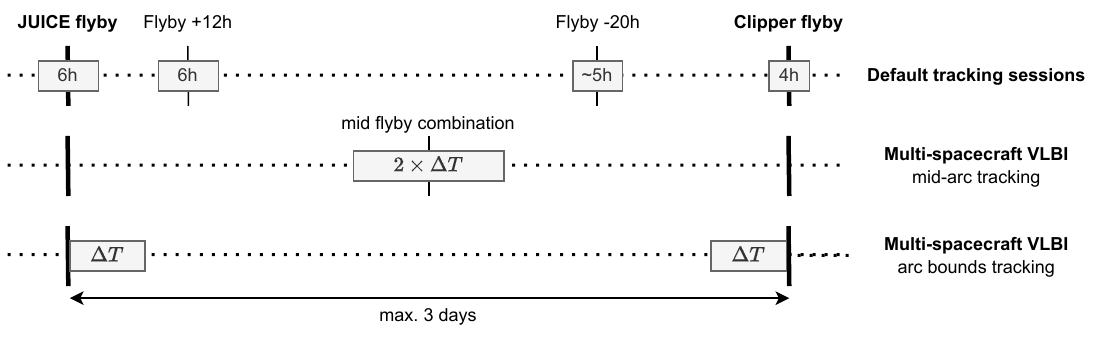}}
	\caption{Multi-spacecraft VLBI tracking configurations, illustrated for a flyby combination where the JUICE flyby occurs before the Europa Clipper one. The grey boxes represent the radio-science tracking sessions, with their durations indicated inside. $\Delta T$ denotes the (varying) duration of the multi-spacecraft tracking arc (see Section \ref{sec:resultsMsVlbi}).}
	\label{fig:msVlbiTrackingOptions}
\end{figure*}

We also tested different tracking scenarios for multi-spacecraft VLBI, essentially distinguishing between two types of configuration (see Figure \ref{fig:msVlbiTrackingOptions}):
\begin{enumerate}
	\item \textit{mid-arc tracking}: single tracking arc centered in-between the JUICE and Europa Clipper flybys involved in the flyby combination of interest; 
	\item \textit{arc bounds tracking}: for each flyby combination, two tracking arcs occurring respectively just after the first close encounter and just before the second one. 
\end{enumerate}
While we varied the duration of the multi-spacecraft tracking arcs, we always ensured that the total tracking duration is identical between the two above cases (i.e. using halved arcs for the arc bounds tracking case). Regarding the quality of the simulated multi-spacecraft VLBI observables, we adopted the two different noise budgets presented in Table \ref{tab:msVlbiErrorBudget}. Finally, navigation Doppler data were also simulated during the longer arcs required for multi-spacecraft tracking, with a noise level of 80 $\mathrm{\mu}$m/s at an integration time of 1h \citep{esocNav}. These Doppler observables were merely included to constrain the empirical accelerations added over these longer arcs (see Section \ref{sec:dynamicalModels}).

\subsection{Estimation strategy }
\label{sec:estimationStrategy}

To quantify the relative improvement of the estimation solution achievable with PRIDE VLBI, we performed multiple covariance analyses, in different scenarios. The covariance matrix $\mathbf{P}$ of the estimated parameters is given by the following \citep{montenbruck2002}:
\begin{align}
	\mathbf{P} = \left(\mathbf{H}^\mathrm{T}\mathbf{W}\mathbf{H} + \mathbf{P}_0^{-1}\right)^{-1}, \label{eq:cov}
\end{align} 
where $\mathbf{W}$ designates the observations weight matrix and $\mathbf{H}$ is the observations partial matrix with respect to the estimated parameters. $\mathbf{P}_0$, on the other hand, contains the a priori covariances of the estimated parameters, accounting for our knowledge of these parameters prior to the estimation. As will be highlighted in Section \ref{sec:parameters}, some of our estimations also include consider parameters (i.e. parameters that are not directly estimated, but whose uncertainties are accounted for in the estimation). The statistical representation of the estimation accuracy is then provided by the so-called consider covariance analysis $\mathbf{P}_c$, defined as:  
\begin{align}
\mathbf{P}_c = \mathbf{P} + \left(\mathbf{P}\mathbf{H}^\mathrm{T}\mathbf{W}\right)\left(\mathbf{H}_c\mathbf{C}\mathbf{H}_c^\mathrm{T}\right)\left(\mathbf{P}\mathbf{H}^\mathbf{T}\mathbf{W}\right)^\mathrm{T}. \label{eq:considerCov}
\end{align}
$\mathbf{P}$, $\mathbf{H}$, and $\mathbf{W}$ refer to the same matrices as in Eq. \ref{eq:cov}, and $\mathbf{H}_c$ and $\mathbf{C}$ respectively designate the observation partials with respect to the consider parameters and the covariance matrix describing our knowledge of these parameters. The formal uncertainties of the estimated parameters are given by the square root of the diagonal elements of $\mathbf{P}$ and $\mathbf{P}_c$. Finally, these formal errors can be propagated to any epoch $t$, the propagated covariance being obtained as follows:
\begin{align}
	\mathbf{P}(t) = \left[\mathbf{\Phi}(t,t_0);\mathbf{S}(t)\right]\mathbf{P}\left[\mathbf{\Phi}(t,t_0);\mathbf{S}(t)\right]^\mathrm{T}, \label{eq:propCov} 
\end{align}
where $\mathbf{\Phi}(t,t_0)$ and $\mathbf{S}(t)$ are the state transition and sensitivity matrices, respectively. Eq. \ref{eq:propCov} can also be applied to propagate the consider covariance $\mathbf{P}_c$ instead of $\mathbf{P}$. 

Covariance analyses, while perfectly adapted for our purposes, inherently rely on a number of simplifying assumptions. In particular, our dynamical models should be able to perfectly represent reality, which is particularly difficult to achieve for the non-conservative accelerations acting on the spacecraft. The resulting formal uncertainties therefore provide a too optimistic statistical representation of the true estimation errors. While we take this into consideration in our discussion, it does not impact the relevance of our approach, since we focus on the \textit{relative} contribution of PRIDE VLBI with respect to a baseline solution. 

However, as discussed in Section \ref{sec:introduction}, modelling inconsistencies do not only yield discrepancies between true and formal errors, but might also complicate the achievement of a consistent, stable solution. Overcoming these issues typically requires an iterative process, starting with reconstructing the spacecraft and flyby moons' orbits locally using the so-called normal points (i.e. arc-wise state solutions for the flybys' central moons, determined at the closest approach). These local estimates of the moons' states can then be reconciled into a global solution in a subsequent step. The main challenges of a direct global estimation of the moons' dynamics will be further discussed in Section \ref{sec:validation}, along with possible mitigation strategies. Because of these foreseen difficulties, we nonetheless chose not to solely focus on a global ephemerides solution, but to also consider the determination of the moons' normal points (i.e. per-flyby solutions) as an intermediate estimation step. In our analyses, we therefore apply both strategies, which are described in more detail in \cite{fayolle2022}:
\begin{itemize}
	\item Local estimation, determining the central moon's normal point for each flyby and each tracking arc during the orbital phase for JUICE. These normal points are estimated perfectly independently from one another \citep[unlike in ][]{fayolle2022}; 
	\item Global estimation, reconstructing a single solution for the moons' orbits over the timelines of the JUICE and Europa Clipper missions. This model has been extended with respect to \cite{fayolle2022} to also account for the concurrent estimation of the central planet's state (see \ref{sec:parameters}). More details on the extended formulation can be found in \ref{appendix:estimation}.
\end{itemize}
In Sections \ref{sec:resultsVlbi} and \ref{sec:resultsMsVlbi}, we thus assess the contribution of PRIDE VLBI data to both types of solution. We also specifically discuss how VLBI could help going from arc-wise state solutions to a single, fully consistent picture of the system's dynamics over the entire missions' timeline (see Section \ref{sec:validation}). 

\subsection{Estimated parameters}\label{sec:parameters}

\begin{table*}[tbp]
	\caption{Estimated and consider parameters included in both our global and local state estimations (see Section \ref{sec:estimationStrategy}). All parameters are by default estimated, except for those reported in italic which are either estimated or included as consider parameters.}
	\label{tab:parameters}
	\centering
	\resizebox{\textwidth}{!}{\begin{tabular}{  l |  c  | c | l  }
        
        \textbf{Parameters} & \textbf{Global estimation} & \textbf{Local estimation} & \textbf{A priori constraint} \\ \hline 

        \multicolumn{4}{l}{}\\
        \multicolumn{4}{l}{\textbf{Jupiter parameters}} \\ \hline 
        
        Initial state & global & not included & 1 km (position) ; 0.1 m/s (velocity) \\
        Gravitational parameter $\mu_0$ & global & global & from Juno, \cite{durante2020}\\
        Zonal gravity coef. up to degree 10 & global & global & from Juno, \cite{durante2020}\\
        Rotation rate and pole orientation & global & global & from Juno, \cite{durante2020}\\
        Inverse tidal quality factor & \multirow{2}{*}{global} & \multirow{2}{*}{not included} & \multirow{2}{*}{none}\\
        at each moon's frequency & & & \\
        \hline

        \multicolumn{4}{l}{}\\
        \multicolumn{4}{l}{\textbf{Moons parameters}} \\ \hline 
        Initial states & global & per arc & 15 km (position) ; 1.0 m/s (velocity)\\
        Gravitational parameters $\mu_i$ & global & global & \cite{schubert2004}\\
        Gravity coef. up to degree and order 2 (Io), & \multirow{2}{*}{global} & \multirow{2}{*}{global} & \cite{schubert2004} for $C_{20}$ and $C_{22}$ \\  
        13 (Europa), 15 (Ganymede), and 9 (Callisto). & & & Kaula's rule\footnotemark{} for other coefficients\\
        Inverse tidal quality factor for each moon & global & not included & none \\
        \hline

        \multicolumn{4}{l}{}\\
        \multicolumn{4}{l}{\textbf{Spacecraft parameters}} \\ \hline
        JUICE and Europa Clipper's states & per arc & per arc & 5 km (position) ; 0.5 m/s (velocity) \\
        Empirical accelerations & per arc & per arc & $5\cdot10^{-8}$ $\mathrm{m}/\mathrm{s}^2$ \\
        \textit{Range biases (for JUICE)} & per pass & per pass & 1.2 m \citep{cappuccio2022} \\
        \textit{Single-spacecraft VLBI biases} & per pass & per pass & calibrator's position uncertainty\\
        \textit{Multi-spacecraft VLBI biases} & per pass & per pass & 0.1 nrad (good) ; 0.25 nrad (poor) \\
        \hline
        			
	\end{tabular}}
\end{table*}

The parameters estimated from the simulated radio-science observables described in Section \ref{sec:simulatedTracking} are reported in Table \ref{tab:parameters}. We distinguish between the global and local state estimation setups introduced in Section \ref{sec:estimationStrategy}, and specify if each parameter is estimated globally or locally (e.g. per arc). The arc and pass definitions refer to those defined in Section \ref{sec:simulatedTracking}. Finally, regarding the moons' gravity field spherical harmonics expansion, we extended it up to the point where expanding it further no longer affects the state estimation results.

The main addition of our baseline setup compared to most radio-science solutions lies in estimating Jupiter's state along with the Galilean moons' orbits. While unnecessary for gravity field analyses, this becomes relevant for moons' ephemerides determination. Existing JUICE and/or Europa Clipper simulations indeed predict extremely low formal uncertainties, reaching sub-metre levels for Ganymede's radial position during JUICE orbital phase \citep[e.g.][]{fayolle2022,magnanini2023}. When facing such accuracy levels, the influence of Jupiter's position error can no longer be neglected and needs to be accounted for in our analyses. 

It should be noted that, in the absence of real data, no unique estimation setup is currently predetermined for JUICE and Europa Clipper radio-science estimations. For this reason, we kept a certain flexibility in our setup. We adopted the configuration in which both range and VLBI biases are estimated as nominal. However, we kept alternative options, such as including observation biases as consider parameters, for additional analyses meant to investigate the sensitivity of our solutions to the estimation setup choice (see Sections \ref{sec:resultsVlbi} and \ref{sec:resultsMsVlbi}). Although estimating observation biases will not necessarily pose a particular challenge, we used these parameters as a proxy to simulate a potential deterioration of the simulation-based solution, possibly too optimistic, when moving to real data analysis. Range biases were selected because of their influence on the determination of the absolute position of the Jovian system, moons, and spacecraft, which directly affects the contribution of PRIDE VLBI measurements. Including biases as consider parameters also allows us to account for the fact that an arc-wise constant value might not be able to adequately model the systematic error in the measurements.  Nonetheless, unless otherwise indicated, the results presented in the rest of this paper are obtained with the nominal setup (i.e., estimating all biases).  

\footnotetext{Kaula's rule: $\sigma=K/l^2$, $K=10^{-5}$, $l$=degree}

\section{Results: single-spacecraft VLBI} \label{sec:resultsVlbi}

This section presents our results regarding the contribution of single-spacecraft PRIDE VLBI measurements of the JUICE spacecraft to the moons' state estimation. We first describe the baseline radio-science solution in Section \ref{sec:baselineSolution}, before presenting the improvement achieved with VLBI for the global and normal points solutions in Sections \ref{sec:resultsVlbiGlobal} and \ref{sec:resultsVlbiNormalPoints}, respectively. The latter addresses the VLBI contribution to the local, intermediate estimation results for the moons' states, which will be essential to eventually achieve the global ephemerides solution discussed in Section \ref{sec:resultsVlbiGlobal}. It is therefore critical to quantify the improvement provided by VLBI with both approaches (see more detailed discussion in Section \ref{sec:validation}). 

\subsection{Baseline solution without VLBI} \label{sec:baselineSolution}

\begin{figure*} [ht!] 
	\centering
	\begin{minipage}[l]{0.49\textwidth}
		\centering
		\subcaptionbox{Io \label{fig:baselineErrorsIo}}
		{\includegraphics[width=1.0\textwidth]{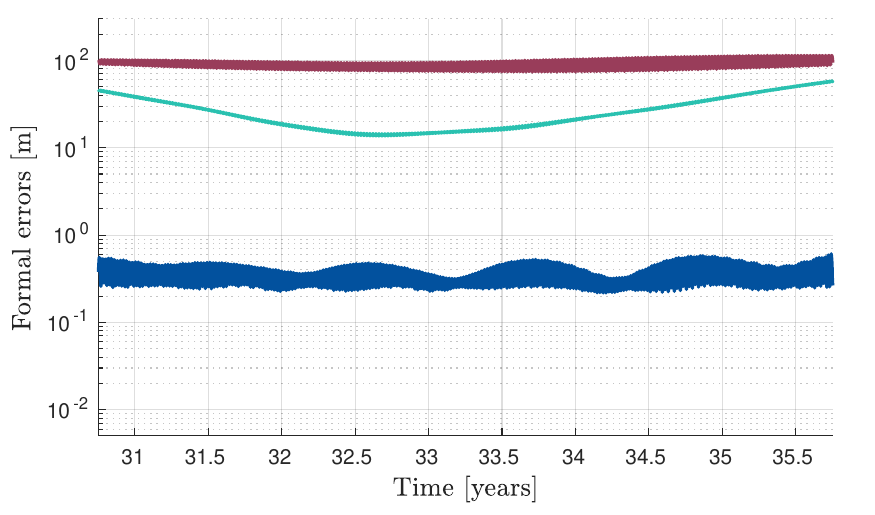}}
	\end{minipage}
	\hfill{} 
	\begin{minipage}[r]{0.49\textwidth}
		\centering
		\subcaptionbox{Europa \label{fig:baselineErrorsEuropa}}
		{\includegraphics[width=1.0\textwidth]{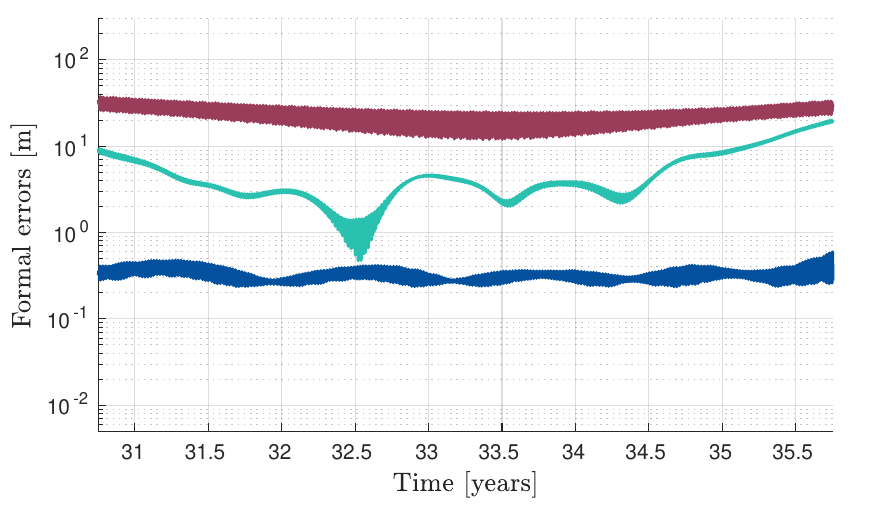}}
	\end{minipage}
	\begin{minipage}[l]{0.49\textwidth}
		\centering
		\subcaptionbox{Ganymede \label{fig:baselineErrorsGanymede}}
		{\includegraphics[width=1.0\textwidth]{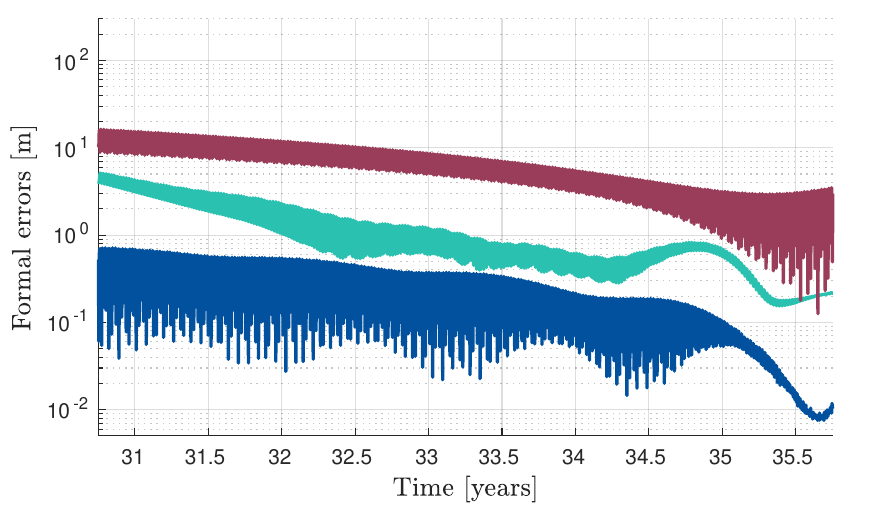}}
	\end{minipage}
	\hfill{} 
	\begin{minipage}[r]{0.49\textwidth}
		\centering
		\subcaptionbox{Callisto \label{fig:baselineErrorsCallisto}}
		{\includegraphics[width=1.0\textwidth]{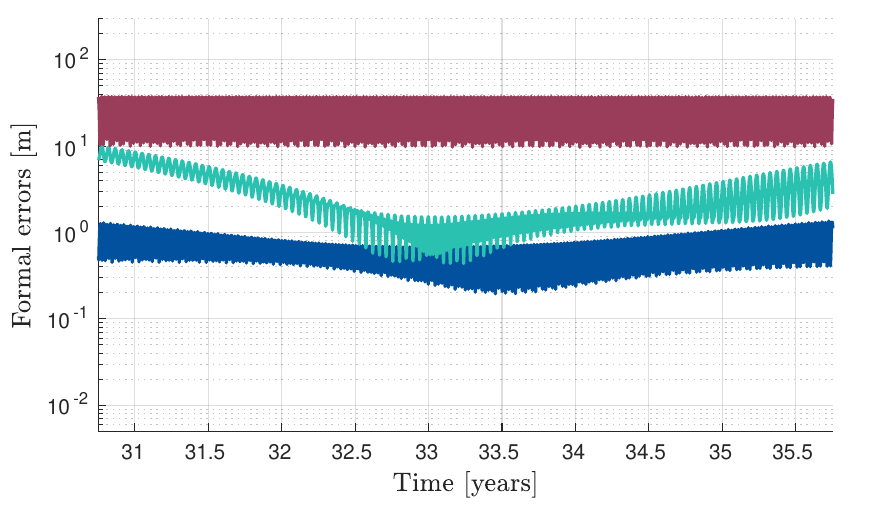}}
	\end{minipage}
    \hfill{}
     \centering
    	\begin{minipage}[r]{0.49\textwidth}
    		\centering
    		\subcaptionbox{Jupiter \label{fig:baselineErrorsJupiter}}
    		{\includegraphics[width=1.0\textwidth]{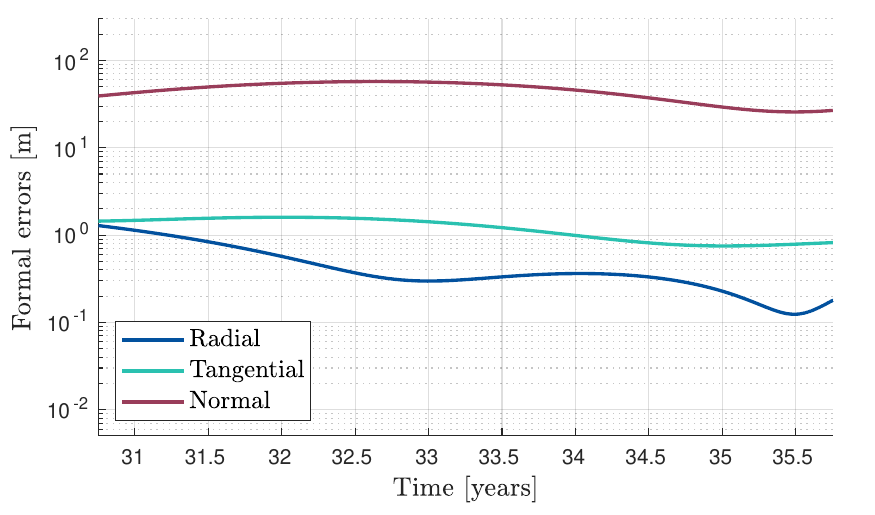}}
	\end{minipage}
	\caption{Baseline formal errors as a function of time for Jupiter and the Galilean moons' positions, estimated from JUICE's and Europa Clipper's range and Doppler simulated data. For the sake of conciseness, the calendar years are shown without preceeding digits "20".}
	\label{fig:baselineErrors}
\end{figure*}

Before quantifying the improvement provided by VLBI, the baseline radio-science solution for the moons' ephemerides, based on JUICE-Europa Clipper range and Doppler (classical) data, must first be briefly discussed. As our analyses focus on PRIDE VLBI specifically, we limit ourselves to a top-level description of the formal uncertainty levels that can be expected from JUICE and Europa Clipper \textit{classical} radio-science measurements. More detailed results and discussions can nonetheless be found in dedicated studies, including more complete analyses of the Jovian system's tidal parameters considering the full effects of tidal dissipation both on the moons' and spacecraft's orbits \citep[][]{cappuccio2020,cappuccio2022,demarchi2021, demarchi2022,magnanini2023,mazarico2023}. 

The formal position uncertainties of our baseline solutions are shown in Figure \ref{fig:baselineErrors}, for both Jupiter and its Galilean satellites. Jupiter's position errors are at the sub-metre level in the radial direction, around 1-2 m for the along-track component and a few tens of metres in the out-of-plane direction. While seemingly small, these errors are not negligible compared to the moons' position uncertainties, confirming the need to account for the Jovian ephemeris error in our estimations (Section \ref{sec:parameters}). These errors are lower than the accuracy of the present Jovian ephemerides if we base ourselves on the differences between existing JPL and INPOP solutions (a few metres to tens of metres, up to 1 km, and up to 4.5 km for the radial, tangential, and normal position components, respectively) \citep[see][]{fayolle2023b}. However, more data from the Juno mission will keep improving these existing solutions, such that the contribution of JUICE and Europa Clipper will be more limited than what our results seem to indicate. It must also be noted that our setup is not intended nor adapted for accurate, global planetary ephemerides determination (outside the scope of our analyses), which would require including all relevant data sets \citep[e.g.,][]{fienga2021} while accounting for more perturbations (main belt asteroid perturbations, refined relativistic models, etc.) over a longer time span.

In the moons' solutions, the flybys and orbital phase also yield very low propagated errors. The signature of the Europa Clipper flybys at Europa, and of JUICE orbital phase around Ganymede, are particularly visible. As expected, the in-plane position components show very low uncertainty levels, on average at the metre level for the radial direction and slightly larger (tens of metres) for the tangential position. This is again a considerable improvement with respect to present solutions, which are accurate to the 10-15 km level (see Table \ref{tab:parameters}). Both the radial and tangential directions are well constrained by Doppler and range data, and the improvement achievable with VLBI is therefore expected to be limited. This, however, does not hold for the normal direction, with larger formal uncertainties at around a hundred metres. Good quality VLBI data (good noise case, see Section \ref{sec:vlbiErrorBudget}) are expected to provide measurements of JUICE's lateral position with an accuracy of about 120-200 m. Considering that multiple VLBI data points will be acquired and that the VLBI geometry is mostly sensitive to the normal direction, this suggests that PRIDE could improve the moons' out-of-plane positions. 

The results presented in Figure \ref{fig:baselineErrors} were obtained with the nominal estimation setup. Nonetheless, a baseline covariance analysis was also conducted with range biases as consider parameters, which increased the position errors by a factor five to eight, depending on the moon and direction, compared to Figure \ref{fig:baselineErrors}. For our analyses, the degradation of the baseline solution's accuracy in the out-of-plane direction, where PRIDE VLBI is expected to provide the largest improvement, is particularly relevant. For our purposes, we nonetheless adopted as \textit{nominal} the setup yielding a more optimistic baseline solution, to avoid overestimating the contribution of PRIDE VLBI observables.

\subsection{VLBI contribution to the global solution} \label{sec:resultsVlbiGlobal}

\begin{table*}[tbp]
	\caption{Improvement in \textbf{averaged} formal position uncertainties (percentage) with respect to the solution obtained with no VLBI, for various VLBI tracking and acquisition scenarios (but VLBI tracking during the flyby phase only). The position errors are computed in the RTN frame, and only improvements larger than 5\% are reported.}
	\label{tab:vlbiGlobal}
	\centering
    \resizebox{0.8\textwidth}{!}{
	\begin{tabular}{ c c | c c c | c c c | c c c | c c c | c c c  }

		\textbf{Cadence} & \textbf{Noise} & \multicolumn{3}{c|}{\textbf{Jupiter}} & \multicolumn{3}{c|}{\textbf{Io}} & \multicolumn{3}{c|}{\textbf{Europa}} & \multicolumn{3}{c|}{\textbf{Ganymede}} & \multicolumn{3}{c}{\textbf{Callisto}} \\
		\textbf{VLBI} & \textbf{budget} & R & T & N & R & T & N & R & T & N & R & T & N & R & T & N \\ 
		\hline
		1 h & poor & - & - & - & - & - & - & - & - & - & - & - & - & - & - & - \\
		1 h & good & - & - & 7.0 & - & - & - & - & - & - & - & - & - & - & - & - \\
		1 h & Ka-band & - & - & 7.6 & - & - & - & - & - & - & - & - & - & - & - & - \\
		\hline
		20 min & poor & - & - & 7.4 & - & - & - & - & - & - & - & - & - & - & - & 5.0 \\
		20 min & good & - & - & 11.5 & - & - & - & - & - & - & - & - & - & 5.1 & 10.1 & 17.1 \\
		20 min & Ka-band & - & 7.8 & 17.5 & - & - & - & - & - & - & - & - & - & 8.8 & 18.5 & 33.8 \\
		\hline
		5 min & poor & - & - & 10.1 & - & - & - & - & - & - & - & - & - & - & 7.2 & 11.8 \\
		5 min & good & - & 8.4 & 18.5 & - & - & - & - & - & - & - & - & - & 9.2 & 19.7 & 36.7 \\
		5 min & Ka-band & 8.7 & 12.9 & 31.4 & - & - & - & - & - & 6.8 & - & - & 7.5 & 15.5 & 28.7 & 55.2 \\
		\hline
		2 min & poor & - & - & 12.4 & - & - & - & - & - & - & - & - & - & 5.2 & 11.7 & 20.7  \\
		2 min & good & 7.0 & 10.9 & 26.6 & - & - & - & - & - & 5.1 & - & - & 5.6 & 13.6 & 26.1 & 50.1 \\
		2 min & Ka-band & 12.5 & 15.2 & 40.9 & - & - & - & - & - & 11.9 & - & 6.8 & 12.5 & 20.1 & 33.4 & 65.1 \\
		\hline
	\end{tabular}}
\end{table*}

After simulating VLBI measurements of the JUICE spacecraft as described in Section \ref{sec:simulatedTracking}, we added these observables to the radio-science estimation. In the following, we describe their contribution to the \textit{global} state solution for the Galilean satellites for different tracking and data quality configurations. The results are summarised in Table \ref{tab:vlbiGlobal}.

We first discuss the GCO phase, during which performing VLBI tracking yields no noticeable improvement. In the best case scenario, the improvement reaches $\sim8\%$ for certain state parameters, but remain around 1-2 \% for most moons' position components. These uncertainty reductions, already negligible, are moreover only achieved in a very optimistic configuration, assuming frequent VLBI tracking sessions (i.e. every week), very dense VLBI outputs (one independent measurement every 2 min) and exceptional data quality (Ka-band noise budget). Consequently, performing VLBI tracking GCO is not worth the negligible improvement it brings to the solution, and we did not consider such tracking options in the rest of our analyses. 

Now focussing on VLBI tracking simulated during the flyby phase, Table \ref{tab:vlbiGlobal} shows the contribution of such observables to the global ephemerides solution for both Jupiter and its moons. The results are expressed as the relative improvement in the propagated position uncertainties with respect to the baseline solution (Figure \ref{fig:baselineErrors}), averaged over the missions' timelines. It must be noted that the error reductions are nearly constant over time, and are therefore adequately represented by the average improvement values provided in Table \ref{tab:vlbiGlobal}. 
Overall, the sensitivity of the PRIDE VLBI contribution to the adopted tracking settings behaves as expected, the improvement becoming stronger with increasing measurement cadence and more accurate observations. In particular, more frequent VLBI data points could notably improve the PRIDE VLBI contribution. However, it is yet unclear if such a high measurement cadence is realistically achievable (due to inter-measurements correlations, see Section \ref{sec:vlbiErrorBudget}). Assessing this will require detailed analyses of the statistical properties of real JUICE VLBI data once available. For our preparatory analyses, we consider a VLBI cadence of one data point every 20 minutes as a reasonable scenario.

As suspected, the improvement is the strongest in the normal direction, for Jupiter and Callisto in particular. The limited number of JUICE flybys (or absence thereof) around Europa and Io implies that only few VLBI data points are strongly sensitive to the dynamics of these two moons, explaining the poor VLBI contribution. For Ganymede, the GCO phase yields an extremely accurate baseline solution, which effectively prevents VLBI tracking from notably improving the solution beyond what Doppler and ranging data can already achieve. Furthermore, even for Jupiter and Callisto, adding JUICE VLBI data only brings limited improvement. Their out-of-plane position errors get reduced by about 11.5\% and 17.1\%, assuming that accurate (good noise case) VLBI measurements can be acquired every 20 minutes.

The lack of a noticeable improvement for the along-track positions of Io, Europa, and Ganymede directly implies that adding VLBI will not further help determine the tidal dissipation in these moons and in Jupiter at these moons' frequencies, as shown by our results in \ref{appendix:tidalDissipation} (Table \ref{tab:baselineErrorsInvQ}). As will be discussed at length in Section \ref{sec:validation}, PRIDE can however facilitate achieving a consistent global solution for the moons' dynamics, which is essential to obtain reliable estimates for tidal dissipation in the Jovian system from JUICE-Europa Clipper radio-science.
It must be noted that the determination of Callisto's in-plane position, mainly in the along-track direction, slightly improves upon adding VLBI measurements. This is an indirect effect of a better determination of Jupiter's tangential (and radial) position achieved with VLBI. This also translates into a small reduction of Callisto's dissipation, as well as Jupiter's dissipation at Callisto's frequency (see Table \ref{tab:baselineErrorsInvQ}). However, the very weak signal of Callisto's dissipation on its own orbit, mostly noticeable in the along-track direction, will still remain far from detectable in JUICE tracking data, even with VLBI. The VLBI contribution to the determination of Jupiter's dissipation at Callisto's frequency also remains limited. While detecting whether Callisto is caught in a resonance locking mechanism could be possible from range and Doppler measurements only, adding VLBI would therefore not be able to have a significant influence on this potential detection.

The above results, which already show very limited VLBI contribution, are furthermore sensitive to the choice of baseline setup and solution. If VLBI biases are not estimated and must be included as consider parameters, adding VLBI measurements actually degrades the moons' state solutions. Remarkably, this still holds when range biases are also treated as consider parameters, i.e. with a more pessimistic baseline solution, which theoretically would leave more improvement margin for PRIDE VLBI. In a consider covariance analysis, the formal errors are automatically raised, by an amount that depends on both the consider parameters covariance and the weights assigned to the observations sensitive to said parameters (see Eq. \ref{eq:considerCov}). For VLBI biases, the very high accuracy of the VLBI observables yields large weights, such that the consider biases significantly affect the covariance results. This highlights the importance of the VLBI calibrators and, more specifically, of their position uncertainty in the ICRF. This could motivate future observation campaigns to identify more suitable or better characterised radio sources, as will be further discussed in the next section.

\subsection{Contribution to local state solutions} \label{sec:resultsVlbiNormalPoints}

\begin{figure*}[ht!]
	\centering
	\makebox[\textwidth][c]{\includegraphics[width=1.0\textwidth]{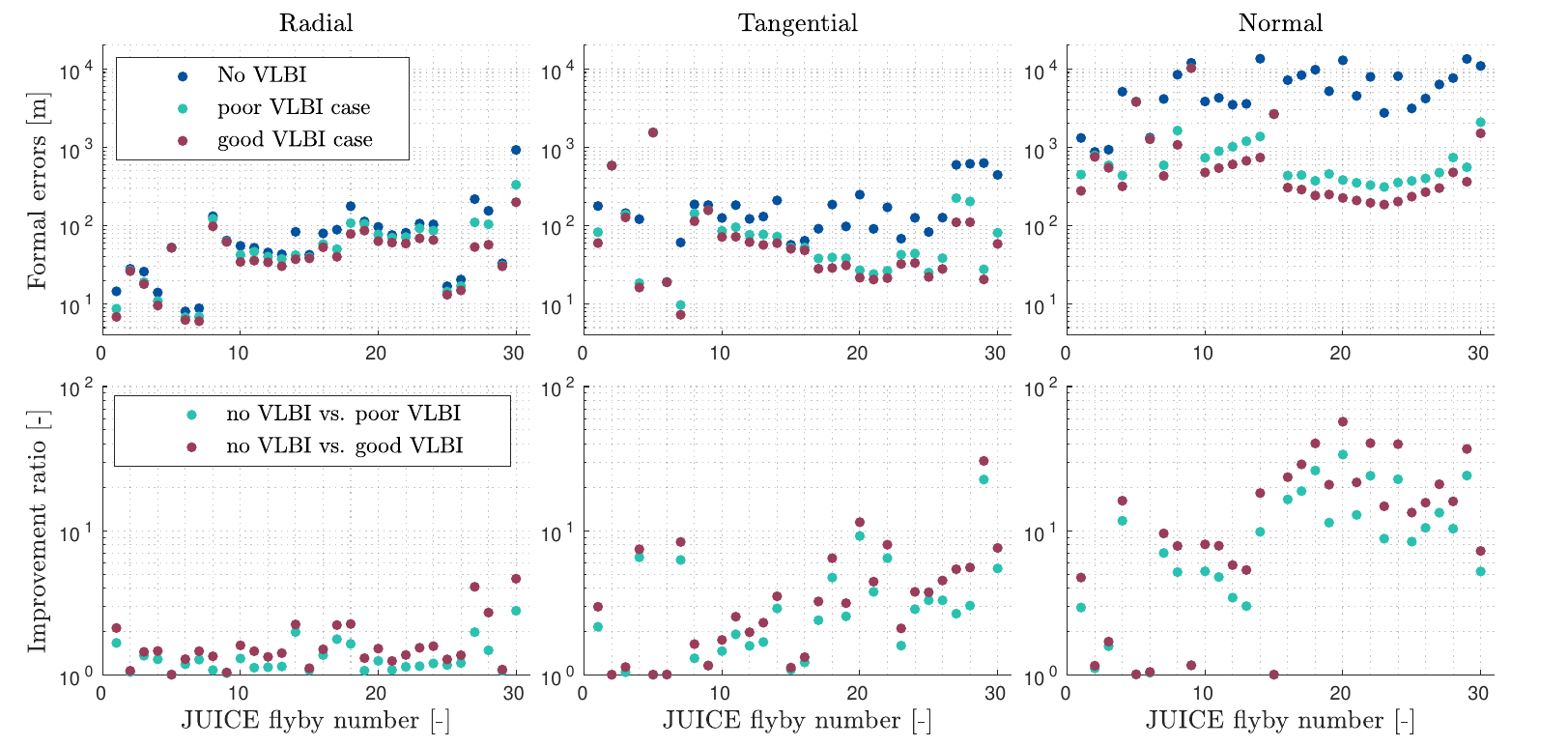}}
	\caption{Formal uncertainties for the normal points generated for each of JUICE's 30 flybys, with and without including VLBI measurements (top panels). The bottom panels show the ratio between the baseline errors (without VLBI) and the uncertainties obtained with VLBI.}
	\label{fig:vlbiNormalPoints}
\end{figure*}

\begin{figure*} [ht!] 
	\centering
	\begin{minipage}[r]{0.49\textwidth}
		\centering
		\subcaptionbox{VLBI biases based on original calibrators (see Figure \ref{fig:calibrators}). \protect\label{fig:vlbiNormalPointsOldBiases}}
		{\includegraphics[width=1.0\textwidth]{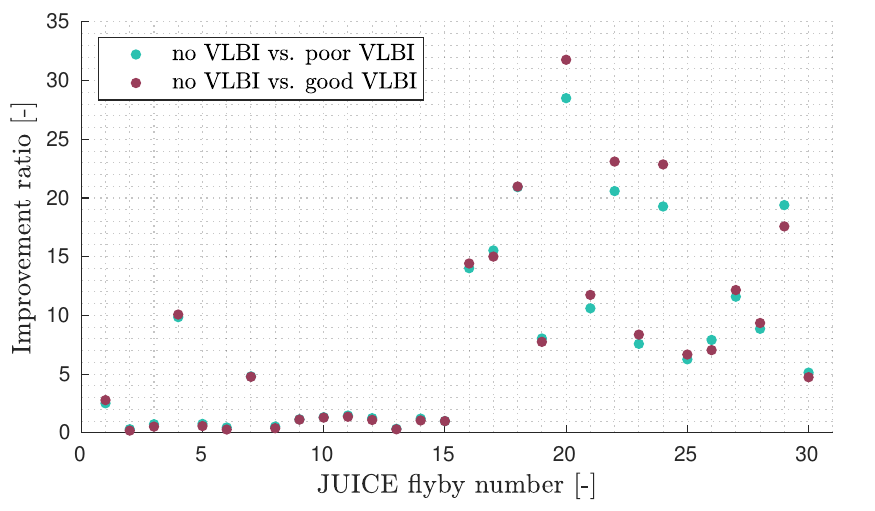}}
	\end{minipage}
	\hfill{}
	\begin{minipage}[l]{0.49\textwidth}
		\centering
		\subcaptionbox{Refined VLBI biases based on artificial, better characterised calibrators. \protect\label{fig:vlbiNormalPointsNewBiases}}
		{\includegraphics[width=1.0\textwidth]{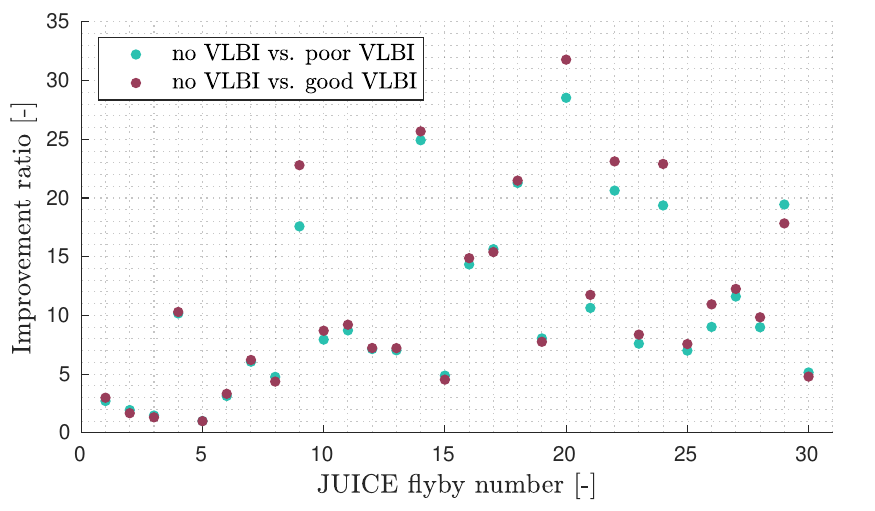}}
	\end{minipage}
	\caption{Improvement ratio of the flyby moon's normal position uncertainty enabled by VLBI tracking, when estimating range biases but including VLBI biases as consider parameters, for different sets of phase calibrators.}
	\label{fig:vlbiNormalPointsBiases}
\end{figure*}

As mentioned in Section \ref{sec:estimationStrategy}, the reconstruction of a global, coherent solution for the Galilean system's dynamics from real data will require proceeding step-by-step, starting with arc-wise state estimations for both the spacecraft and the moons. The results in Section \ref{sec:resultsVlbiGlobal} indicate that PRIDE VLBI data will have a very limited influence on the quality of the final ephemerides. In this section, we however assess how much PRIDE VLBI will contribute to the process of achieving such a global solution, by quantifying the improvement of the moons' normal points achievable with VLBI data (see more detailed discussion in Section \ref{sec:validation}). The moons' local state uncertainties are actually significantly larger than those achieved with a global estimation. Each flyby is indeed processed independently, without constraining the local solutions for a given moon to form a consistent, single trajectory \citep{fayolle2022}. The VLBI contribution to the moons' normal points is therefore stronger than for the global estimation, for which the extreme accuracy of the baseline solution limits the margin for further improvement (see Section \ref{sec:resultsVlbiGlobal}). 

Figure \ref{fig:vlbiNormalPoints} shows the position formal errors obtained for each normal point (per-flyby solution, see Section \ref{sec:estimationStrategy}), with and without VLBI, as well as the improvement ratio between the two solutions. As observed in the global estimation results (Section \ref{sec:resultsVlbiGlobal}), VLBI tracking mostly reduces the flyby moon's position in the out-of-plane direction. The VLBI contribution to the flyby moon's normal position only appears negligible for a few flybys, namely flybys 2, 3, 5, 6, 9, and 15. They actually correspond to situations where VLBI measurements cannot be simulated, either because no calibrator is available (see Figure \ref{fig:calibrators}) or because the tracking visibility conditions are not met (elevation lower than 15 deg or signal occulted by another moon, or by Jupiter). 
For the rest of the JUICE flybys, however, VLBI data significantly reduce the normal point uncertainties in the out-of-plane direction, with an averaged improvement ratio of about 10 and 20 for the poor and good VLBI error budgets, respectively. VLBI tracking can also help refine the flyby moon's along-track position, depending on the flyby geometry and accuracy of the baseline solution. 
Unlike for the moons, the improvement in the spacecraft's local state at each flyby's closest approach is however very limited, reaching at most 30\% with the best VLBI noise budget and in the normal direction only.

Focussing on the PRIDE contribution to the moons' state solutions, the good VLBI noise case automatically yields lower uncertainties than the more pessimistic error budget. Nonetheless, the latter can still provide a significant improvement with respect to the baseline solution (see Figure \ref{fig:vlbiNormalPoints}, especially for the normal direction). This demonstrates the potential of PRIDE VLBI data as a powerful means to refine our local estimation of the moons' states, irrespective of the measurements accuracy. Section \ref{sec:validation} will further explore the key role that these more accurate normal points can play in helping us reconstruct a consistent global solution for the moons' dynamics. 

Similarly as for the global estimation case (Section \ref{sec:resultsVlbiGlobal}), we investigated the sensitivity of our normal point results with respect to the choice of estimation setup. To this end, we re-conducted our analysis while assuming that VLBI biases cannot be estimated and must be accounted for as consider parameters. As expected, this weakens the contribution of PRIDE VLBI to the local flyby moon's states. Figure \ref{fig:vlbiNormalPointsOldBiases} shows the improvement ratio of the flyby moon's normal position uncertainty achieved with VLBI in such an estimation setup (as in Figure \ref{fig:vlbiNormalPoints}, the other two directions show much more limited improvement). In addition to the few above-mentioned flybys with unfavourable VLBI tracking conditions, flybys 8 to 14 now also show negligible improvement. These flybys overlap with the period of the JUICE Jovian tour when the identified phase calibrators are characterised by abnormally poorly constrained ICRF positions (referred to as poor calibrators in the following). Such calibrators yield large systematic VLBI errors for flybys 8 to 15 (Figure \ref{fig:calibrators}), which strongly affect the estimation if they cannot be better determined. 

We therefore assessed how better-suited calibrators (which could be found with a dedicated campaign, e.g. \citealt{duev2016}) would improve the determination of the corresponding normal points. To this end, we substituted the poor calibrators with an artificial one, with a more typical position uncertainty. The latter was set to the average value computed among all suitable calibrators identified over JUICE flyby phase (i.e. all calibrators with a position error lower than 2 nrad in Figure \ref{fig:calibrators}). Figure \ref{fig:vlbiNormalPointsNewBiases} shows how this indeed further improves the solution achieved with VLBI data for flybys 8-15. This further demonstrates the importance of using adequate, and sufficiently characterised radio source as phase calibrators, to avoid introducing systematic errors in our estimation and to maximise the added-value of VLBI tracking.

\section{Results: multi-spacecraft in-beam VLBI} \label{sec:resultsMsVlbi}

This section presents the solution improvement achievable by performing multi-spacecraft VLBI tracking of the JUICE and Europa Clipper spacecraft. Using the 11 flyby combinations identified in Section \ref{sec:msVlbi}, we simulated these unique observables and included them in our state estimation. Sections \ref{sec:resultsMsVlbiGlobal} and \ref{sec:resultsMsVlbiNormalPoints} respectively discuss the contribution of multi-spacecraft VLBI to the global ephemerides solution and normal points estimation. We used the same baseline solution as presented in Section \ref{sec:baselineSolution}.

\subsection{VLBI contribution to the global solution} \label{sec:resultsMsVlbiGlobal}

\begin{table*}[tbp]
	\caption{Improvement in \textbf{averaged} formal position uncertainties (percentage) with respect to the solution obtained with no VLBI, for various multi-spacecraft VLBI tracking scenarios (see Section \ref{sec:simulatedTracking} and Figure \ref{fig:msVlbiTrackingOptions}). The position errors are computed in the RTN frame, and only improvements larger than 5\% are reported.}
	\label{tab:msVlbiGlobal}
	\centering
	\resizebox{\textwidth}{!}{\begin{tabular}{  c  c  | c | c c c | c c c | c c c | c c c | c c c  }
		
			\multicolumn{2}{c|}{\multirow{2}{*}{\textbf{Tracking arc}}} & \textbf{Noise} & \multicolumn{3}{c|}{\textbf{Jupiter}} & \multicolumn{3}{c|}{\textbf{Io}} & \multicolumn{3}{c|}{\textbf{Europa}} & \multicolumn{3}{c|}{\textbf{Ganymede}} & \multicolumn{3}{c}{\textbf{Callisto}} \\
			
			\multicolumn{2}{c|}{} & \textbf{budget} & R & T & N & R & T & N & R & T & N & R & T & N & R & T & N \\ 
			\hline
			
			$2\times$4h & arc bounds & poor & 7.8 & 5.9 & 41.2 & - & - & - & 6.9 & - & 18.8 & - & 5.1 & - & 7.4 & 7.1 & 7.0 \\ 
			$2\times$4h & mid-arc & poor & 9.2 & 6.0 & 39.6 & 10.0 & 7.3 & 7.9 & 7.1 & 6.3 & 17.9 & 6.2 & 6.4 & 7.1 & 8.1 & 7.2 & 6.5 \\ \hline
			
			$2\times$8h & arc bounds & poor & 9.9 & 7.5 & 44.1 & 	8.6 & 7.8 & 8.4 & 8.9 & 7.6 & 22.5 & 6.6 & 7.1 & 9.4 & 9.4 & 9.3 & 9.8 \\ 
			$2\times$8h & mid-arc & poor & 10.7 & 7.2 & 42.1 & 12.8 & 10.6 & 13.5 & 9.3 & 10.1 & 26.0 & 7.4 & 8.3 & 11.1 & 9.4 & 9.4 & 9.5 \\ \hline
			
			$2\times$12h & arc bounds & poor & 10.1 & 6.9 & 45.1 & 11.4 & 11.5 & 8.9 & 10.7 & 10.0 & 27.9 & 7.2 & 8.5 & 11.4 & 9.6 & 10.4 & 11.2 \\
			$2\times$12h & mid-arc & poor & 12.3 & 8.0 & 43.7 & 12.2 & 12.4 & 8.1 & 10.7 & 10.9 & 32.5 & 8.3 & 9.7 & 15.1 & 10.1 & 10.8 & 11.4 \\ \hline	
			
			$2\times$24h & arc bounds & poor & 14.0 & 10.5 & 50.7 & 14.6 & 14.6 & 10.9 & 13.3 & 12.2 & 34.2 & 10.2 & 11.0 & 15.6 & 13.2 & 16.2 & 20.6 \\ 
			$2\times$24h & mid-arc & poor & 16.5 & 10.0 & 48.6 & 13.9 & 14.8 & 9.7 & 13.8 & 11.7 & 44.4 & 12.5 & 13.2 & 24.3 & 13.4 & 15.6 & 17.9 \\ \hline	
			
			\multicolumn{2}{c|}{\textbf{Full tracking}} & \textbf{poor} & \textbf{18.7} & \textbf{11.9} & \textbf{51.2} & \textbf{16.1} & \textbf{16.6} & \textbf{12.2} & \textbf{16.3} & \textbf{13.6} & \textbf{47.4} & \textbf{15.0} & \textbf{15.4} & \textbf{28.4} & \textbf{15.7} & \textbf{18.0} & \textbf{21.5} \\ 
				
			\multicolumn{17}{c}{} \\ \hline
			
			$2\times$4h & arc bounds & good &13.6 & 10.3 & 55.7 & 6.0 & 7.9 & - & 10.3 & 6.3 & 31.5 & 7.9 & 8.9 & 10.6 & 11.2 & 10.1 & 9.7 \\
			$2\times$4h & mid-arc & good & 16.0 & 10.6 & 56.1 & 12.3 & 10.8 & 10.1 & 10.0 & 8.6 & 30.5 & 10.3 & 10.4 & 13.4 & 13.8 & 13.1 & 13.2 \\ \hline
			
			$2\times$8h & arc bounds & good & 16.1 & 13.4 & 58.0 & 12.4 & 12.3 & 10.9 & 12.3 & 10.5 & 35.9 & 10.1 & 10.9 & 14.2 & 13.5 & 15.1 & 18.8 \\
			$2\times$8h & mid-arc & good & 18.8 & 13.6 & 58.9 & 15.2 & 14.5 & 15.4 & 12.9 & 12.7 & 40.3 & 12.7 & 13.2 & 19.5 & 15.2 & 16.1 & 18.4 \\ \hline
			
			$2\times$12h & arc bounds & good & 18.3 & 16.3 & 60.4 & 14.9 & 15.6 & 11.8 & 15.3 & 12.8 & 42.8 & 11.9 & 12.8 & 20.4 & 16.1 & 18.4 & 24.3 \\ 
			$2\times$12h & mid-arc & good & 20.7 & 13.9 & 60.5 & 14.8 & 16.2 & 11.3 & 14.2 & 12.6 & 46.7 & 14.6 & 15.3 & 25.3 & 16.4 & 18.8 & 22.0 \\ \hline
			
			$2\times$24h & arc bounds & good & 23.8 & 18.8 & 65.9 & 17.7 & 18.8 & 13.7 & 18.4 & 15.1 & 50.4 & 16.9 & 16.6 & 28.3 & 20.0 & 26.2 & 37.2 \\
			$2\times$24h & mid-arc & good & 26.3 & 18.4 & 65.3 & 17.1 & 19.0 & 13.8 & 18.6 & 14.7 & 56.6 & 21.3 & 19.8 & 38.2 & 20.2 & 25.3 & 34.0 \\ \hline
			
			\multicolumn{2}{c|}{\textbf{Full tracking}} & \textbf{good} & \textbf{28.5} & \textbf{20.1} & \textbf{67.6} & \textbf{19.2} & \textbf{21.1} & \textbf{15.2} & \textbf{21.3} & \textbf{16.4} & \textbf{59.1} & \textbf{24.5} & \textbf{22.4} & \textbf{41.5} & \textbf{21.8} & \textbf{27.4} & \textbf{37.5} \\ 	
	\end{tabular}}
\end{table*}

Table \ref{tab:msVlbiGlobal} presents the relative improvement of the moons' global state solutions achieved with multi-spacecraft tracking VLBI, for the different tracking configurations defined in Section \ref{sec:simulatedTracking}. As with single-spacecraft VLBI, we only provide the average improvement, given that PRIDE VLBI contribution to the propagated errors is almost constant over the missions' duration. As expected, longer tracking arcs and more accurate measurements strengthen the contribution of the multi-spacecraft VLBI observables to the solution. In the following, we adopt tracking arcs of $2\times$8 hours as the nominal configuration. Longer tracking arcs are deemed too optimistic regarding the additional tracking resources that would be required both onboard the spacecraft and on ground. The full tracking configuration covering the entire time gap separating the JUICE and Europa Clipper flybys thus depicts an optimal, yet practically unrealistic tracking scenario, but is merely intended to quantify the strongest improvement possibly achievable. 

On average, for identical tracking durations, it first appears more beneficial to acquire multi-spacecraft VLBI measurements in the mid-arc tracking scenario. However, the nominal transmitting sessions are centered around the flybys (Section \ref{sec:simulatedTracking}, Figure \ref{fig:msVlbiTrackingOptions}).  For combinations with a rather long time gap between the two flybys ($>$ 1 day), mid-arc tracking might thus require planning a full additional tracking session in-between the two flybys, for both spacecraft, with the necessary resource allocations that this implies. The arc bounds tracking strategy, on the other hand, will exploit the fact that each spacecraft is already transmitting close to its flyby. This approach effectively limits the additional tracking resources with respect to the mid-arc option. Comparing the results of the $2\times8$ h arc bounds and $2\times4$ h mid-arc tracking cases in Table \ref{tab:msVlbiGlobal}, which require comparable extra resources, we recommend adopting the arc bounds strategy when planning future multi-spacecraft VLBI tracking. 

Overall, the contribution of multi-spacecraft VLBI measurements to the moons' global states is stronger than for single-spacecraft VLBI (see Section \ref{sec:resultsVlbiGlobal}). This also holds when comparing the results obtained with poor multi-spacecraft and good single-spacecraft VLBI, despite them sharing comparable noise budgets (Tables \ref{tab:errorBudget} and \ref{tab:msVlbiErrorBudget}). 
The improvement is particularly strong for Jupiter and Europa. For Europa, the reason for this significant improvement is twofold. First, Europa is involved in 7 out of the 11 flyby combinations during which multi-spacecraft VLBI tracking is performed (Table \ref{tab:msVlbiOptions}). Second, most of these Europa flybys are Europa Clipper flybys, and Europa Clipper's coarser radio-science solution leaves more margin for improvement compared to JUICE's. The contribution to Jupiter's state estimation, on the other hand, is an indirect effect of the measurement geometry: by constraining the relative positions of the two spacecraft close to some of their flybys, multi-spacecraft VLBI constrains the moons' relative dynamics in their orbit around Jupiter, which greatly helps refine Jupiter's position. 

Unlike in the single-spacecraft VLBI case, our baseline setup, by estimating both range and VLBI biases, yields a rather conservative quantification of the multi-spacecraft VLBI contribution. Systematic VLBI errors are indeed small (Section \ref{sec:msVlbi}) and therefore do not strongly affect the solution. However, the improvement attainable with VLBI only gets larger when using a slightly more pessimistic baseline solution such as the one obtained when range biases are not estimated (see \ref{appendix:msVlbiDifferentSetups}). The above strengthens the robustness of our findings, hinting that multi-spacecraft VLBI measurements might improve the moons' ephemerides solution further than suggested by Table \ref{tab:msVlbiGlobal}, depending on the quality of the baseline solution. 

It must moreover be noted that the time elapsed in-between the two flybys is shorter than one day for 3 combinations out of 11 in Table \ref{tab:msVlbiOptions}. For these combinations, multi-spacecraft tracking could be performed without extending the nominal tracking sessions (Section \ref{sec:simulatedTracking}). Interestingly, the solution improvement achievable with these three combinations is not negligible, as shown in Table \ref{tab:msVlbiFlybysCombinations} in \ref{appendix:msVlbiFlybysCombinations}. In the good VLBI  noise case, Jupiter and Callisto's normal position errors still get reduced by about 53\% and 27\%, respectively, against 58\% and 36\% with all 11 flyby combinations. This result demonstrates that non-negligible improvement could still be achieved with multi-spacecraft VLBI without necessarily requiring extra resources.

Finally, we investigated the role played by the navigation Doppler data simulated during the multi-spacecraft VLBI tracking arcs (see Section \ref{sec:simulatedTracking}). However, they only contributed to estimating the extra empirical accelerations added to our setup to account for the perturbations (e.g. manoeuvres) influencing the spacecraft's orbits over longer arcs. No improvement of the moons' orbit solutions was indeed noticed when adding Doppler navigation data only. The uncertainty reductions reported in Table \ref{tab:msVlbiGlobal} can therefore be confidently attributed to multi-spacecraft VLBI.

\subsection{VLBI contribution to local state solutions} \label{sec:resultsMsVlbiNormalPoints}

\begin{figure*}[ht!]
	\centering
	\makebox[\textwidth][c]{\includegraphics[width=1.0\textwidth]{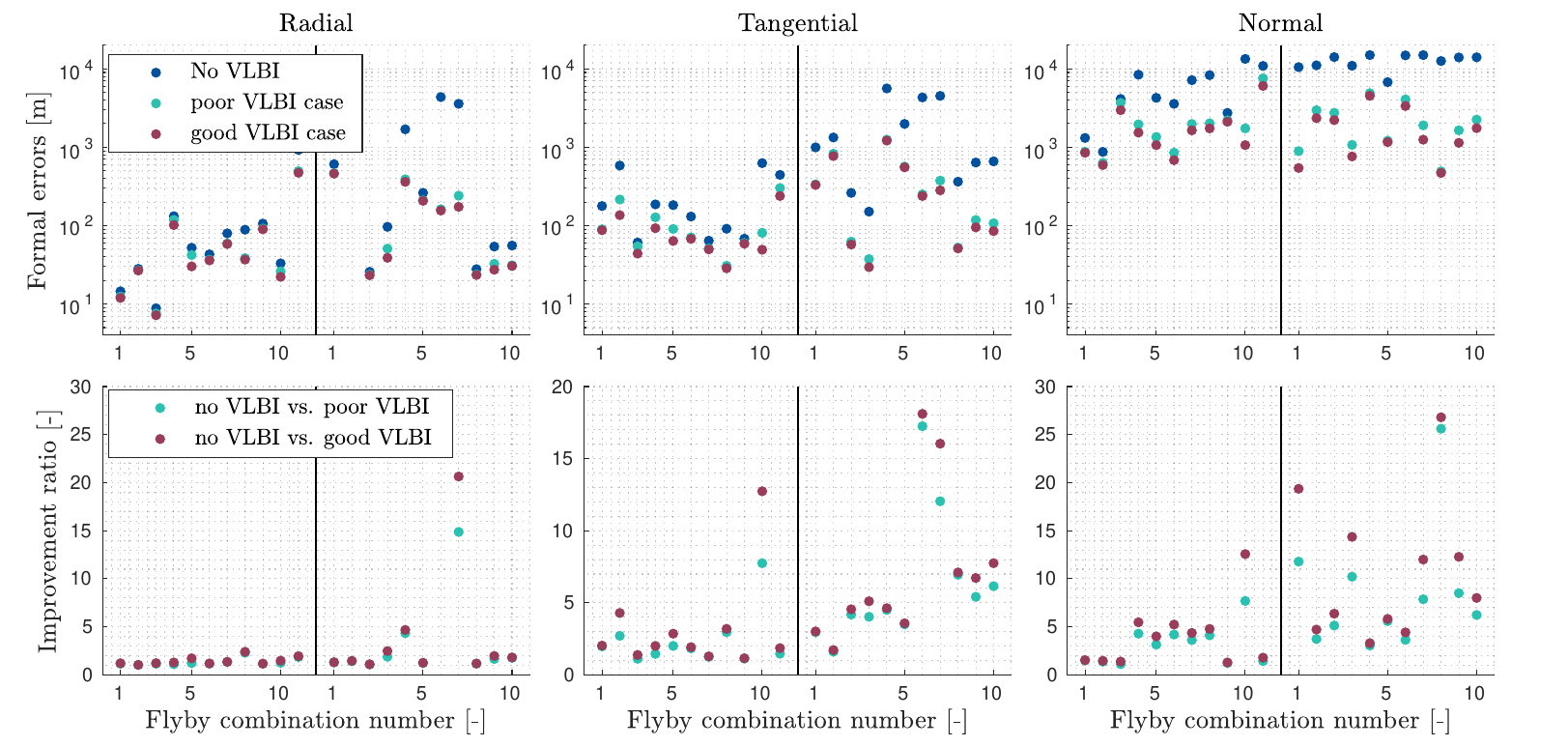}}
	\caption{Formal uncertainties for the normal points generated for each of the 11 flyby combinations in Table \ref{tab:msVlbiOptions}, with and without including multi-spacecraft VLBI measurements (top panels). The bottom panels show the ratio between the baseline errors (without VLBI) and the uncertainties obtained with VLBI. The black vertical lines distinguish between the results for the 11 JUICE flybys (left hand side) and the corresponding 11 Europa Clipper flybys (right hand side).}
	\label{fig:msVlbiNormalPoints}
\end{figure*}

\begin{table*}[ht!]
	\caption{Improvement ratio for the flyby moons' position uncertainties (percentage), when adding different VLBI data sets with respect to the baseline estimation (without VLBI). The values reported in this table are averaged over the 11 flyby combinations during which multi-spacecraft tracking is possible (Table \ref{tab:msVlbiOptions}).}
	\label{tab:allVlbiNormalPoints}
	\centering
    \small
    \resizebox{0.8\textwidth}{!}{
	\begin{tabular}{ c | c | c | c c c }
		
		\textbf{} & 	\multicolumn{1}{|c|}{\multirow{2}{*}{\textbf{Noise budget}}} & \multicolumn{1}{|c|}{\multirow{2}{*}{\textbf{VLBI data set}}} & \multicolumn{3}{|c}{\textbf{Improvement ratio}} \\
		& & & R & T & N \\ 
		\hline
		
		\multicolumn{1}{c|}{\multirow{3}{*}{JUICE flybys}} & \multicolumn{1}{|c|}{\multirow{3}{*}{poor VLBI}} & single-spacecraft VLBI & 1.4 & 4.3 & 8.9 \\
		 & & multi-spacecraft VLBI & 1.4 & 2.3 & 11.3 \\
		 & & both VLBI & 1.7 & 8.4 & 17.6 \\ \hline
		 
		 \multicolumn{1}{c|}{\multirow{3}{*}{JUICE flybys}} & \multicolumn{1}{c|}{\multirow{3}{*}{good VLBI}} & single-spacecraft VLBI & 1.8 & 5.8 & 13.4 \\
		 & & multi-spacecraft VLBI & 1.5 & 3.2 & 4.0 \\
		 & & both VLBI & 2.3 & 8.4 & 17.6 \\
		 \hline 
		 	 
		 \multicolumn{1}{c|}{\multirow{3}{*}{Europa Clipper flybys}} & \multicolumn{1}{|c|}{\multirow{3}{*}{poor VLBI}} & single-spaeccraft VLBI & - & - & - \\
		 & & multi-spacecraft VLBI & 5.3 & 6.2 & 8.3 \\
		 & & both VLBI & 5.4 & 8.2 & 17.3 \\ \hline
		 
		 \multicolumn{1}{c|}{\multirow{3}{*}{Europa Clipper flybys}} & \multicolumn{1}{c|}{\multirow{3}{*}{good VLBI}} & single-spacecraft VLBI & - & - & - \\
		 & & multi-spacecraft VLBI & 6.0 & 7.1 & 10.7 \\
		 & & both VLBI & 6.2 & 9.5 & 23.6 \\
		\hline
	\end{tabular}}
\end{table*}

Multi-spacecraft VLBI measurements were also included in the normal points determination, to assess the contribution of such observables to the moons' arc-wise solutions. Figure \ref{fig:msVlbiNormalPoints} shows the improvement achieved for the two moons involved in each flyby combination. While the contribution of multi-spacecraft VLBI is again the largest for the moons' normal positions, the improvement in this direction is lower than with single-spacecraft VLBI (see Figures \ref{fig:vlbiNormalPoints} and \ref{fig:msVlbiNormalPoints}). On the contrary, however, the reduction of the in-plane position uncertainties is slightly stronger with multi-spacecraft VLBI. On average, the moons' radial and along-track local positions indeed get reduced by more than a factor three and four, respectively, even with the poor VLBI error budget. 

This can be explained by the difference in the nature of the observables: while single-spacecraft VLBI provides a direct measure of JUICE's lateral position in the ICRF, multi-spacecraft observations are only sensitive to JUICE and Europa Clipper relative position. They therefore indirectly constrain the relative motion of the flybys' moons, instead of their absolute positions. Consequently, depending on the geometry of the flyby combination, the signature of the moons' out-of-plane positions in the multi-spacecraft VLBI observables is not systematically as strong as it would be for single-spacecraft VLBI. On the other hand, multi-spacecraft tracking might provide slightly tighter constraints on the moons' in-plane motion. This strong dependency on the tracking geometry also explains the variability of the multi-spacecraft VLBI contribution from one flyby combination to another (see Figure \ref{fig:msVlbiNormalPoints}). 

Interestingly, the improvement is much stronger for the central moons of Europa Clipper's flybys than for JUICE's. This logically follows from Europa Clipper's baseline state estimation being less accurate, due to the lower quality of Europa Clipper's tracking (see Section \ref{sec:simulatedTracking}). The uncertainties of the moon's normal points however become comparable between JUICE and Europa Clipper's flybys once multi-spacecraft VLBI is included. Starting from a coarser solution, the relative improvement is thus stronger for the Europa Clipper estimation.

Finally, it should be mentioned that the improvement provided by multi-spacecraft VLBI is not limited to the flyby moons' state solutions, but also extends to the spacecraft orbit determination. The absolute formal errors and subsequent VLBI improvement vary from flyby to flyby due to their different geometries. While it is outside the scope of this paper to provide a detailed analysis of spacecraft orbit determination results, we can still make general observations. Overall, adding multi-spacecraft VLBI brings the spacecraft's along-track and normal position errors to similar levels (several tens of metres), while radial uncertainties amount to a few metres. Again, the exact improvement that this represents with respect to the baseline solution depends on the flyby. Overall, it is nonetheless stronger in the out-of-plane direction: when taking the average over the 11 flyby combinations, multi-spacecraft tracking can lower the spacecraft's position uncertainties at closest approach by about a factor 5, 2.5, and 50 in the radial, tangential, and normal directions, respectively. This improvement is moreover rather independent of the choice of VLBI noise budget. While not the primary focus of our analyses, smaller uncertainty ellipses for the spacecraft's local states might greatly help disentangling mismodelling effects affecting either the spacecraft's or the moons' dynamics, as will be further discussed in Section \ref{sec:validation}.

Overall, the general trends highlighted in Figure \ref{fig:msVlbiNormalPoints} for a specific case (8h-long arcs, arc bounds tracking) do not strongly depend on the tracking configuration considered. All tracking setups reported in Table \ref{tab:msVlbiGlobal} for the global estimation were also tested for the normal points determination. 
Interestingly, unlike what was observed in Section \ref{sec:resultsVlbi}, simulating multi-spacecraft VLBI tracking close to both flybys, and not in the middle of the arc, yields better results. This can be expected when reconstructing local state solutions at flyby time: the moon's dynamical signature is stronger in tracking measurements acquired immediately before or after the close encounter. 

Finally, we also quantified the combined improvement attainable when both single- and multi-spacecraft VLBI observables are included in the estimation. The improvement ratio of the flyby moons' position components with single-spacecraft VLBI only, multi-spacecraft VLBI only, and both types of VLBI are reported in Table \ref{tab:allVlbiNormalPoints}. Adding all VLBI measurements does significantly reduce the normal points' uncertainties for the flyby moons, in all three directions. Given that PRIDE is a JUICE experiment, no single-spacecraft VLBI tracking was considered for the Europa Clipper spacecraft. Remarkably, however, the solution for Europa Clipper's flyby moons notably improves when adding JUICE's nominal tracking measurements to the estimation. This is an indirect effect of the better state solution achieved for JUICE's flyby which, via the constraints provided by the multi-spacecraft tracking measurements, also constrain Europa Clipper's flyby solution. In addition to these quantitative improvements, the synergy between single- and multi-spacecraft VLBI reaches its full potential when exploited to validate the baseline radio-science solution(s), as will be explored in Section \ref{sec:validation}.

\section{PRIDE VLBI as a powerful validation means} \label{sec:validation}

While our results indicate that PRIDE VLBI may not significantly contribute to the moons' global state estimation (Sections \ref{sec:resultsVlbiGlobal} and \ref{sec:resultsMsVlbiGlobal}), it can greatly reduce \textit{local} state estimation uncertainties and play a key role in helping us eventually achieve a global solution. As discussed in Section \ref{sec:estimationStrategy}, when reconstructing the dynamics of natural satellites from spacecraft tracking, mismodelling of the spacecraft or moons' dynamics might impede the direct reconstruction of a global solution for the moons' orbits. A global state estimation for the moons indeed requires the spacecraft and moons' dynamical models to be consistent over both short and long timescales (typical flyby duration, i.e. a few hours, vs. entire mission). In particular, combining all available flybys at a given moon in a single solution increases the observation timespan, such that additional perturbations and possibly mismodelled effects become relevant. As mentioned in Section \ref{sec:introduction}, such modelling issues led to solution instabilities and prevented the reconstruction of a global ephemeris for Titan and Dione from Cassini flybys' radio-science in \cite{durante2019} and \cite{zannoni2020}, respectively. While a solution was eventually achieved for Titan by \cite{lainey2020}, the Cassini example perfectly illustrates the difficulties that one can expect for future JUICE-Europa Clipper analyses: we will need to proceed gradually from local estimation alternatives \citep[e.g.,][]{durante2019} to a robust global solution \citep{lainey2020}.

For JUICE and Europa Clipper radio-science analyses, this modelling consistency requirement is even made more severe by the very good accuracy levels for the moons' ephemerides predicted by simulations \citep{fayolle2022,magnanini2023}. For these formal uncertainties to be physically meaningful, our dynamical models should be consistent to the same (sub-meter) level. For the spacecraft's dynamics, this makes the coherent modelling of all spacecraft perturbations essential (manoeuvres, solar radiation pressure, accelerometer errors, etc.). Based on past Cassini data analyses, issues related to specific aspects of the moons' dynamical models will also arise. In particular, the modelling of (frequent-dependent) tidal dissipation in the central planet and the moons, as well as variations of the central planet's gravity field and rotation, are expected to be critical \citep{durante2019,zannoni2020}.  

Traditionally, the moons' orbits are first solved for in an arc-wise manner, using the normal points approach mentioned in Section \ref{sec:estimationStrategy}. This first step is also essential to the determination of a robust and accurate spacecraft orbit solution. This strategy is moreover perfectly adapted to gravity field studies \citep{durante2019}, with the added benefit of circumventing the above-mentioned modelling challenges. It indeed relaxes the modelling requirements by letting the moon's \textit{local} state solution absorb part of the models' inaccuracies \citep[see][for a detailed discussion]{fayolle2022}. When reconstructing the moons' dynamics using a decoupled approach, these normal points (i.e. arc-wise state estimates and their corresponding formal uncertainties) are then used as observables to reconstruct a global solution \citep[see more detail in][]{fayolle2022}. Generating per-flyby, \textit{local} state solutions for the moons will therefore be an indispensable first step when determining the Galilean moons' ephemerides from JUICE-Europa Clipper radio-science. These local state estimations will be the groundwork for gradually progressing towards a global, coupled inversion of the spacecraft and moons' dynamics over the entire mission(s) duration. Local solutions can also help assessing the global solution's quality once such a solution is attained, as will be further discussed in the following.

By providing an additional set of independent measurements of the spacecraft's lateral position in the ICRF, PRIDE VLBI not only yields an improved local estimation, but can also help us moving from the normal points determination to the reconstruction of a single, consistent solution for the moons' orbits. In the following, we propose an iterative PRIDE-based validation strategy, showing how VLBI data can improve and/or validate the estimation solutions at various stages of this process. Special attention is paid to the assessment of the statistical realism and robustness of the solution, essential to its accurate interpretation. Section \ref{sec:validationLocal} first discusses how the refined normal points obtained with VLBI (Sections \ref{sec:resultsVlbiNormalPoints} and \ref{sec:resultsMsVlbiNormalPoints}) can be used to detect possible inconsistencies in our models and investigate their possible causes. Capitalising on these local analyses, Section \ref{sec:validationGlobal} then presents several validation steps exploiting PRIDE VLBI to facilitate the estimation of a global, coupled solution for the moons' dynamics.

\subsection{Application to the local state estimations} \label{sec:validationLocal}

\begin{figure*}[ht!]
	\centering
	\makebox[\textwidth][c]{\includegraphics[width=0.8\textwidth]{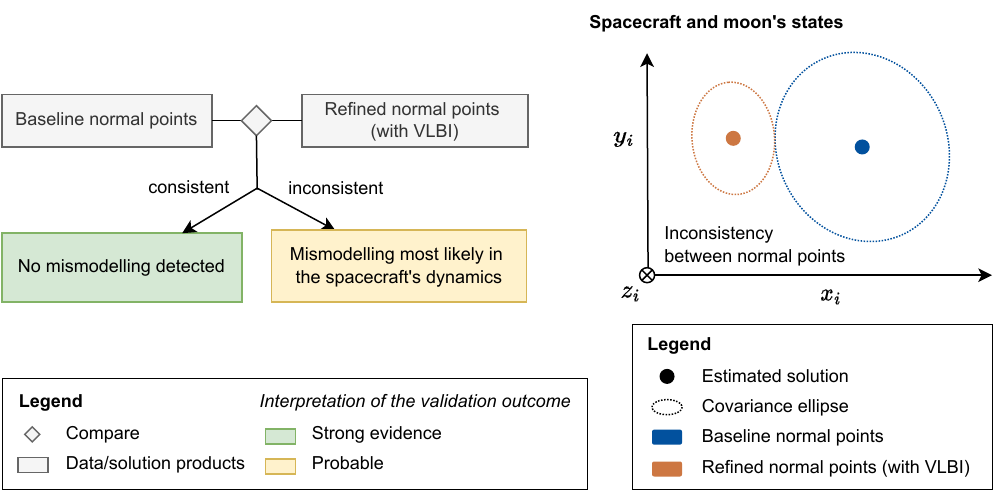}}
	\caption{Validation of the statistical consistency between the moons' normal points reconstructed with and without VLBI measurements.}
	\label{fig:validation_1}
\end{figure*}

As shown by our results presented in Sections \ref{sec:resultsVlbiNormalPoints} and \ref{sec:resultsMsVlbiNormalPoints}, adding VLBI measurements to the estimation can significantly lower the state uncertainties associated with the moons' normal points. In addition to this promising quantitative improvement, verifying the statistical consistency of the arc-wise state solutions obtained with and without VLBI (Figure \ref{fig:validation_1}) can bring valuable insights into the consistency of our models, which might later affect our ability to achieve a global solution, or the robustness of said solution.

Detecting inconsistencies between the normal points with and without VLBI would suggest either dynamical mismodelling issues or larger-than-expected systematic VLBI errors. The latter is nonetheless rather unlikely, considering that we will have a good estimate of the expected error budget of our VLBI measurements. Moreover, we should be able to identify such observation errors by analysing our post-fit residuals. In particular, they should manifest themselves as non-flat, incompressible residuals for VLBI observations specifically, not observable for Doppler and/or range measurements. Finally, it must be noted that large systematic errors in the VLBI measurements would most likely be caused by the use of poor calibrators, which can clearly identified (see Figure \ref{fig:calibrators}). For the flybys that would show inconsistencies between the normal points with and without VLBI, a detailed characterisation campaign of the radio source used as calibrator can be performed a posteriori \citep[see e.g.][]{duev2012}. This would yield a better phase calibration, and will allow us to eliminate unexpectedly large systematic biases from our measurements.  

After addressing VLBI measurement-related issues, remaining inconsistencies between the refined and nominal normal points (i.e. without and with VLBI, respectively, as illustrated by the right-hand side of Figure \ref{fig:validation_2}) can be safely attributed to dynamical modelling issues. Given that the reconstruction of the flybys' normal points does not force the moons' local states to form a single, consistent trajectory, modelling inconsistencies are, at this stage, more likely to originate from the spacecraft's dynamics.

\subsection{Application to the global state estimation} \label{sec:validationGlobal}

Following the careful analyses of our local solutions described in Section \ref{sec:validationLocal}, the next step is to perform a global, coupled estimation of the moons' dynamics. While such an estimation strategy is expected to yield the most statistically consistent state solution \citep{fayolle2022}, successfully achieving the above will require proceeding gradually. Provided that instabilities caused by modelling inconsistencies do not prevent us from obtaining such a global solution, modelling errors are still expected to translate into large, non-flat residuals and/or large true-to-formal errors.    
A fully statistically consistent and robust ephemerides solution for the Galilean moons from classical radio-science measurements thus cannot be achieved directly, but can only be attained through an iterative process. This will imply, in particular, detecting and overcoming various inconsistencies and inaccuracies in our models affecting the quality and realism of the solution. In the following, we explore how PRIDE VLBI can facilitate this process, by identifying, isolating, and whenever possible mitigating potential modelling inconsistencies. In the subsequent discussion, we designate by \textit{preliminary} global solution an intermediate global estimation result (without VLBI), obtained when working towards a final, fully consistent solution. This solution corresponds to the nominal estimation setup described in Section \ref{sec:parameters}.

\subsubsection{VLBI as independent measurements}

\begin{figure*}[ht!]
	\centering
	\makebox[\textwidth][c]{\includegraphics[width=0.8\textwidth]{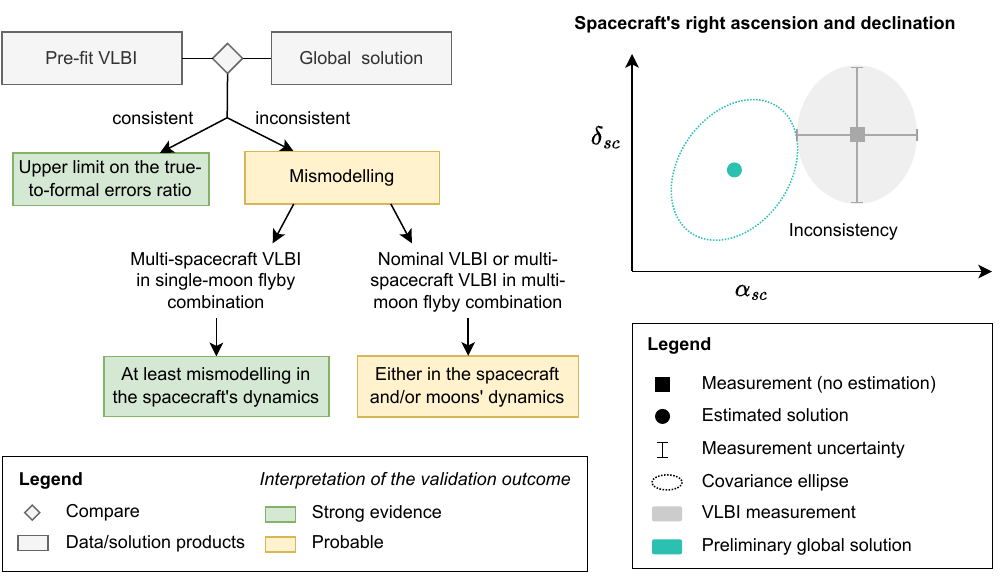}}
	\caption{Comparison between a preliminary global solution (without VLBI) and the pre-fit VLBI measurements to detect or quantify possible inconsistencies in the estimation.}
	\label{fig:validation_2}
\end{figure*}

\begin{figure*}[ht!]
	\centering
	\makebox[\textwidth][c]{\includegraphics[width=1.0\textwidth]{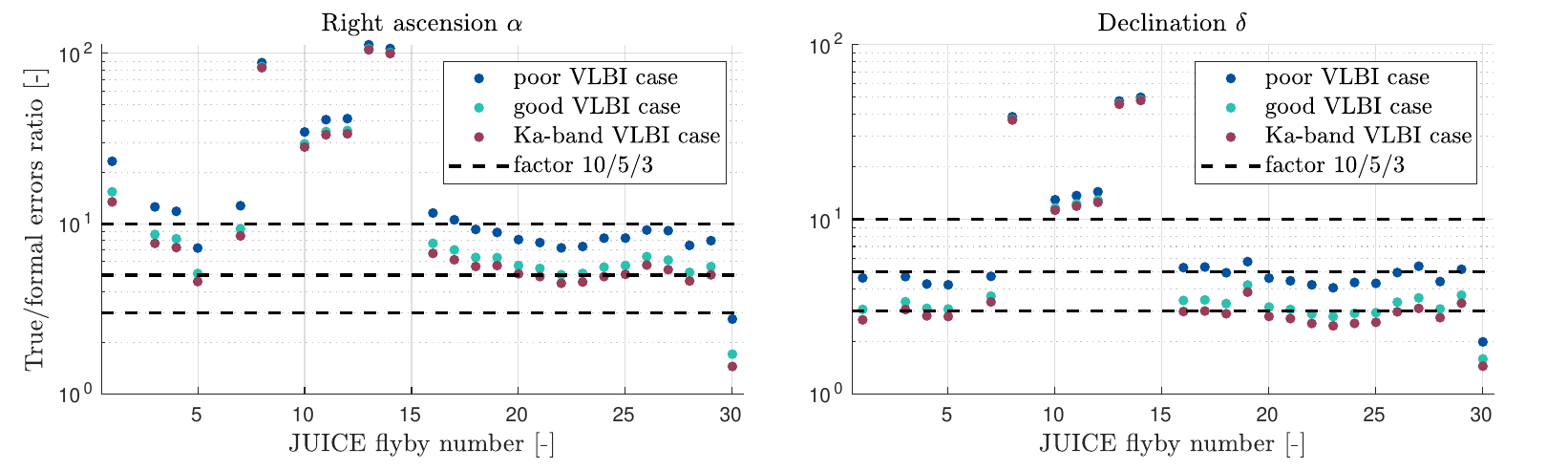}}
	\caption{Detection threshold for possible inconsistencies between the global solution (without VLBI) and the single-spacecraft pre-fit VLBI measurements, expressed as the minimum true-to-formal error ratio required for such discrepancies to be detectable.}
	\label{fig:validationVlbi}
\end{figure*}

\begin{figure*}[ht!]
	\centering
	\makebox[\textwidth][c]{\includegraphics[width=1.0\textwidth]{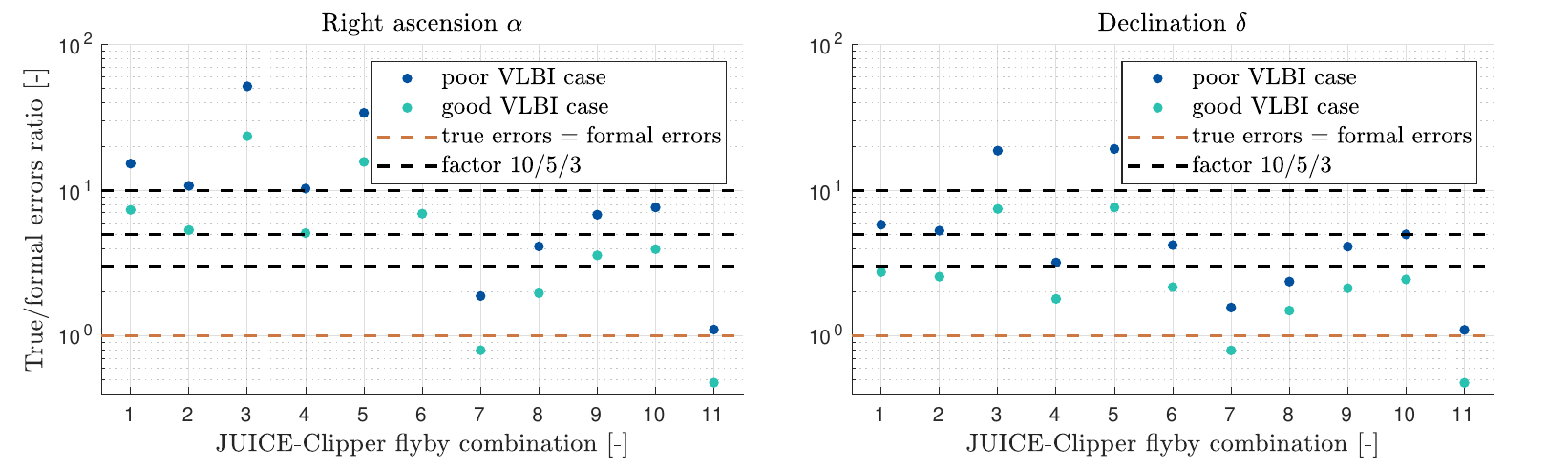}}
	\caption{Detection threshold for possible inconsistencies between the global solution (without VLBI) and the multi-spacecraft pre-fit VLBI measurements, expressed as the minimum true-to-formal error ratio required for such discrepancies to be detectable.}
	\label{fig:validationMsVlbi}
\end{figure*}

An important first validation step to assess the statistical realism of the \textit{preliminary} global solution is to verify that said solution is compatible with the VLBI measurements of the spacecraft's angular position. As illustrated in Figure \ref{fig:validation_2}, the pre-fit VLBI (i.e. \textit{raw} measurements, as in not included in the estimation) should fall within the error ellipse defined by the preliminary global solution's covariance.
As the VLBI data are not yet included in the estimation, they are only affected by the measurement error, and not by potential dynamical mismodelling. A discrepancy between the VLBI measurements and the preliminary global solution would thus indicate either an unquantified systematic bias in the VLBI data (see Section \ref{sec:validationLocal}), or issues in the global estimation (e.g. large true-to-formal errors ratio). 

Relatively large true-to-formal error ratios can be expected when reconstructing natural satellites' ephemerides from radio-science, compared to astrometry-based solutions. The observational constraints on the moons' dynamics, derived from spacecraft tracking measurements, are indeed indirect in nature. Estimations of physical parameters from tracking data are thus affected by modelling inacccuracies in the spacecraft's dynamics, and therefore typically show larger true-to-formal error ratios. Based on previous radio-science estimations of e.g. natural bodies' gravity fields and rotations, ratios of about 10 can be expected \citep[e.g.][]{milani2001, konopliv2011,mazarico2015}. Last but not least, as previously mentioned, determining the moons' orbits from radio-science imposes to consistently model the dynamics of both the spacecraft and moons, a requirement even more stringent for JUICE and Europa Clipper analyses due to the expected low formal uncertainties \citep{fayolle2022}. In our analyses, we thus considered three, five, and ten as a realistic range of true-to-formal errors ratios. Any detection threshold comparable to or lower than these ratios indicates that VLBI tracking might be realistically sensitive to possible inconsistencies in the preliminary global solution, or at least provide an upper limit on the true errors for the moons' states.

To quantify the probability that VLBI data can detect discrepancies between true and formal errors, we projected the error ellipse of the spacecraft's position given by the preliminary global solution onto the plane-of-the-sky, to be compared with the expected VLBI measurement uncertainty. As illustrated in Figure \ref{fig:validation_2}, we could then determine the minimum true estimation errors in JUICE's right ascension and declination for the global solution \textbf{not to overlap} with the VLBI measurement. In practice, a discrepancy can be detected if the estimated solution, within the confidence region statistically described by its formal uncertainties, is not consistent with the VLBI measurement, even when accounting for the uncertainty of the latter. This imposes a limit on the \textit{minimum} true estimation error required for the VLBI observable to detect a possible inconsistency, referred to as the discrepancy detection threshold in the following. 

We computed this threshold for each flyby, both for single- and multi-spacecraft VLBI tracking (Figures \ref{fig:validationVlbi} and \ref{fig:validationMsVlbi}, respectively). For the former, we compare the uncertainty in JUICE's right ascension and declination predicted by the global estimation (without VLBI) with the expected VLBI accuracy. The use of multi-spacecraft VLBI for validation is nearly identical, except that we focus on the \textit{relative} lateral position of JUICE and Europa Clipper with respect to each other. Figures \ref{fig:validationVlbi} and \ref{fig:validationMsVlbi} show, for each flyby or flyby combination, the ratio between the true and formal errors in the spacecraft's right ascension and declination corresponding to the discrepancy detection threshold defined above. A threshold value equivalent to a realistic true-to-formal errors ratio for our analyses (see discussion above) indicates that VLBI measurements can be meaningfully used to investigate possible inconsistencies in the estimation.

Figure \ref{fig:validationVlbi} shows that single-spacecraft VLBI tracking could meaningfully contribute to validating the preliminary global solution for most flybys. Assuming the worst VLBI error budget and a true-to-formal errors ratio equal to ten, PRIDE VLBI could detect inconsistencies in JUICE's right ascension for 14 out of 30 flybys. For JUICE's declination, the discrepancy detection threshold is lower, and an estimation error only five times larger than the formal uncertainty would be detectable for half of the flybys. Improving the VLBI precision would lower this threshold further: in the good error budget case, VLBI data would be sensitive to any true-to-formal error ratio larger than three in declination for flybys 20 to 30. Interestingly, the detection level is rather consistent between the different flybys, with the exception of flybys 8-15 with poor VLBI calibrators (see Figure \ref{fig:calibrators} and discussion in Section \ref{sec:resultsVlbi}).

Figure \ref{fig:validationMsVlbi} highlights similar validation opportunities for multi-spacecraft VLBI tracking, for the 11 possible flyby combinations identified in Section \ref{sec:msVlbi}. As in the single-spacecraft VLBI case, inconsistencies in declination will be easier to detect: a true-to-formal error ratio of three should be detectable for 9 out of 11 combinations, in the good VLBI error case. The slightly larger variability of the detection threshold compared to Figure \ref{fig:validationVlbi} can be explained by the relative nature of multi-spacecraft VLBI tracking: how the accuracy of such measurements compares to the preliminary solution depends on the relative geometry of JUICE and Europa Clipper, and on whether part of the preliminary solution's uncertainties cancel out when computing the error ellipse for the two spacecraft's relative right ascension and declination.  

Overall, our results show that pre-fit VLBI will be able to detect inconsistencies in the preliminary global solution for a number of flybys and/or flyby combinations, provided that the true errors are large enough with respect to the formal uncertainties. Alternatively, detecting no discrepancy would demonstrate the realism of the estimation solution, and allow us to put an upper limit on the true-to-formal error ratio. 

In most cases, the validation step described above will however not be sufficient to precisely identify the source of the potential inconsistencies, if detected.
A notable exception, highlighted in Figure \ref{fig:validation_2}, arises for multi-spacecraft VLBI acquired during a single-moon flyby combination (flyby combinations 3, 5, 7, and 8, see Table \ref{tab:msVlbiOptions}). In such a tracking configuration, the VLBI data points are almost insensitive to the moon's state estimation, except for the possible slight change in the moon's position error during the time elapsed between the JUICE and Europa Clipper flybys. This effect, however, is deemed small, especially for flybys combination 3 in which only three hours separate the two flybys around Europa. The outcome of our first validation step for these single-moon flyby combinations will therefore primarily depend on the consistency of the spacecraft's orbit solution. As such, they will represent a unique opportunity to isolate modelling issues affecting the spacecraft's dynamics.

\subsubsection{Comparing local and global solutions}

\begin{figure*}[tbp!]
	\centering
	\makebox[\textwidth][c]{\includegraphics[width=0.8\textwidth]{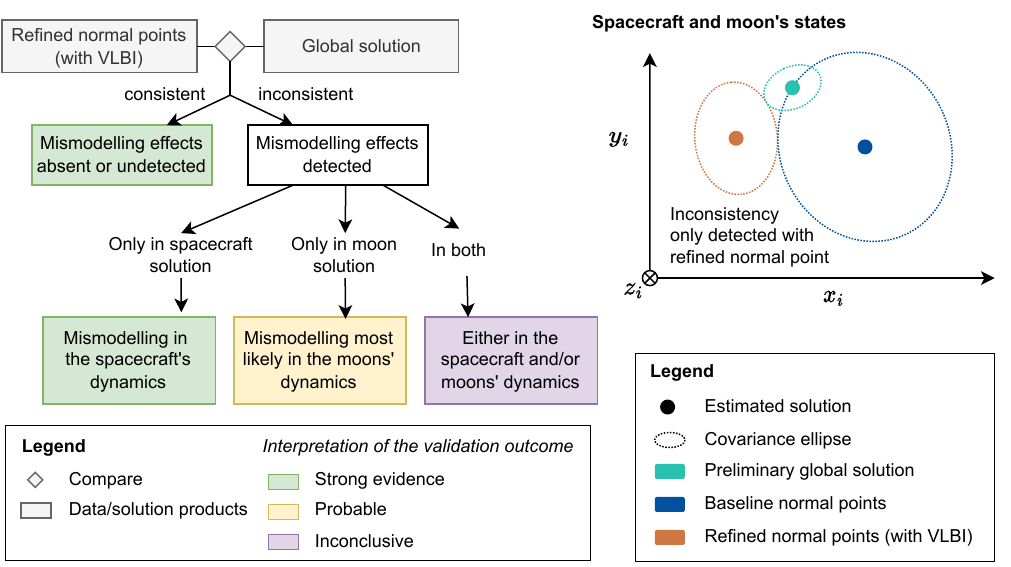}}
	\caption{Validation strategy exploiting the moons' refined normal points obtained with VLBI to detect possible inconsistency in a global preliminary solution.}
	\label{fig:validation_3}
\end{figure*}

Finally, the state solution provided by the global estimation at the time of a flyby $i$ should be statistically compatible with the corresponding normal point. As shown in Sections \ref{sec:resultsVlbiNormalPoints} and \ref{sec:resultsMsVlbiNormalPoints}, VLBI tracking, either in single or multi-spacecraft configuration, can significantly reduce the uncertainty ellipses of the moons' normal points. This enhances the potential of this local vs. global state estimation comparison, by facilitating the detection of possible inconsistencies. The refined arc-wise solutions, with reduced uncertainties, indeed become sensitive to much smaller discrepancies (see Figure \ref{fig:validation_3}).

As in the previous validation steps, the main challenge is to identify the source of the observed discrepancies. Our ability to do so will strongly depend on the state parameters concerned by the said inconsistencies (see Figure \ref{fig:validation_3}). The arc-wise state solutions for the spacecraft and the moons can be analysed separately to try disentangling different mismodelled effects. If inconsistencies are only detected in the spacecraft's state solution, they are more likely to originate from mismodelling of the spacecraft dynamics, while the opposite is true for the moons' solution. However, no firm conclusion can be drawn if the discrepancies concern both the spacecraft and moons' solutions. 

Critically, the outcomes of this validation step must be considered in light of previous results. As described in Section \ref{sec:validationLocal}, we should be able to eventually discriminate between VLBI systematic errors and dynamical mismodelling effects. Furthermore, combining the different tests described above (Figures \ref{fig:validation_2} and \ref{fig:validation_3}) will help further isolate modelling issues specifically affecting the spacecraft or moons' dynamics. We will moreover be able to confirm our conclusions by exploiting the unique potential of multi-spacecraft VLBI in the four single-moon flyby combinations identified in Table \ref{tab:msVlbiOptions}, as such measurements will be robust against errors in the moons' state solutions.

\section{Conclusion} \label{sec:conclusion}

Building on the previous work by \cite{dirkx2017}, we investigated the contribution of PRIDE VLBI to the Galilean moons' ephemerides solution, in the context of the JUICE and Europa Clipper missions. We considered both a global and local state estimation, the latter representing a necessary intermediate step to eventually achieve a coherent solution for the moons' dynamics over the entire missions' timeline. We simulated \textit{single-spacecraft} VLBI measurements of the JUICE spacecraft, but also explored the possibility to perform simultaneous VLBI tracking of JUICE and Europa Clipper (multi-spacecraft VLBI). We quantified the contribution of both types of VLBI data to the moons' global and local state estimations, under various tracking and data quality scenarios. 

Compared to the initial analysis by \cite{dirkx2017} and as highlighted in Section \ref{sec:introduction}, the analysis presented in this paper more rigorously accounts for all couplings between the Galilean moons and spacecraft's dynamics, and also considers the effect of the Jovian ephemeris error on the moons' global solution. However, moving to a joint JUICE-Europa Clipper setup (not relevant at the time of the study by \citealt{dirkx2017}) and to the newest JUICE trajectory significantly improves the quality of the solution achievable with range and Doppler only. This notably reduces the margin for further improvement attainable with VLBI, compared to the first results by \cite{dirkx2017}, where a significant improvement was obtained for the out-of-plane positions of Ganymede and Callisto in particular.

With our updated setup, both single- and multi-spacecraft VLBI measurements no longer significantly improve the \textit{global} ephemerides solution for the Galilean moons, the contribution of the latter nonetheless being stronger. For realistic tracking configurations, the improvement provided by single- and multi-spacecraft VLBI can reach up to 17\% (for Callisto) and 36\% (for Europa), respectively, assuming good VLBI data quality. The attainable improvement is severely limited by the very accurate baseline solution already achieved with range and Doppler data.

It must be noted that our single-spacecraft VLBI results proved rather sensitive to systematic errors in the VLBI measurements. For each tracking pass, the position error of the selected phase calibrator can thus have a significant influence, as highlighted in Figure \ref{fig:vlbiNormalPointsBiases}. This could, however, be mitigated in various ways. Our results indeed motivate future campaigns to densify the phase reference calibrators currently identified within the required patch of the sky, or to refine our knowledge of the ICRF position of known calibrators. In particular, we identified a specific period, overlapping with 8 out of 30 JUICE flybys, during which finding better calibrators would be critical to performing high-quality PRIDE observations (see Section \ref{sec:calibrators}). The lack of suitable calibrators in Ka-band also calls for dedicated reference source densification campaigns. Alternatively, one could exploit the fact that some tracking arcs rely on the same calibrator, as extracting a common bias over several arcs will be easier. It might moreover be possible to reduce VLBI errors by using multiple visible phase calibrators during a single pass. However, properly assessing both the feasibility and actual potential of such a strategy would require dedicated further analyses.

The possible contribution of PRIDE VLBI is moreover not limited to a quantitative improvement of the state estimation. For each flyby, the local estimation of the central moon's state (i.e. normal point) represents an essential step before a global, fully consistent solution can be reconstructed from all flybys combined. The contribution of PRIDE VLBI to the flyby moons' normal points is much stronger than for the global ephemerides solution. This is most noticeable in the out-of-plane direction where poor-quality single- and multi-spacecraft VLBI data respectively reduce the position uncertainty by a factor 10 and 6 on average. This highlights the crucial role that PRIDE VLBI can play in the progression towards a global solution for the moons' dynamics.

VLBI also offers multiple opportunities to validate and improve the statistical realism of the baseline solution derived from classical radio-science observables. To exploit this potential, we have designed a PRIDE VLBI-based validation plan, which exploits two features of the VLBI data set. First, PRIDE provides \textit{independent} measurements, which the baseline solution can be compared against. Second, the ability of VLBI data to reduce the moons' local state uncertainties will facilitate the detection of possible inconsistencies in the estimation. In particular, the careful analysis of the observation residuals and state estimation solutions in different configurations will help disentangle inconsistency sources, from observation errors to various dynamical modelling discrepancies. The unique geometry of multi-spacecraft tracking VLBI data acquired when both JUICE and Europa Clipper are performing a flyby around the same moon will be particularly valuable to isolate specific mismodelling issues. 

PRIDE VLBI will therefore greatly contribute to overcoming dynamical modelling issues in the estimation, gradually working towards the very low uncertainty levels predicted by simulations for the moons' ephemerides and the Jovian system's tidal dissipation parameters \citep[e.g.][]{fayolle2023,magnanini2023}. As such, PRIDE will play an indirect, yet crucial, role in the reconstruction of an unprecedentedly accurate and fully consistent solution for the Galilean moons' dynamics, essential to further our understanding of the Galilean system's long-term evolution.

\section*{Acknowledgments}

The authors acknowledge the use of Astrogeo Center database maintained by Leonid Petrov. This research was partially funded by ESA’s OSIP (Open Space Innovation Platform) program (M.F.), and supported by CNES, focused on JUICE (V.L.). MF would like to express their gratitude to Andrea Minervino Amodio for the very insightful discussions and valuable inputs on the use of PRIDE VLBI measurements for validation purposes.

\bibliographystyle{apalike} 
\bibliography{references.bib}

\appendix

\setcounter{figure}{0}
\setcounter{table}{0}

\section{Concurrent state estimation of a central planet and its satellite(s) from several orbiting spacecraft} \label{appendix:estimation}

As mentioned in Section \ref{sec:estimationStrategy}, our global estimation setup follows the coupled model described in \cite{fayolle2022}. We recall that the states of the moons are then determined globally, while the spacecraft's dynamics are solved for in an arc-wise manner. In this appendix, we expand the mathematical formulation provided in \cite{fayolle2022} to include the central planet (here Jupiter) in the estimation and to account for several spacecraft (here JUICE and Europa Clipper). 

In this more complete configuration, additional implementation subtleties arise when handling the various central body dependencies. The different bodies and spacecraft's states are indeed typically expressed and estimated with respect to their central body, which might be included in the propagation. The full state vector is defined as
\begin{align}
	 \mathbf{y}(t) = 
	\begin{pmatrix}
	\mathbf{y}_{_{P}}(t) \\
	\mathbf{y}_{_{M}}(t) \\
	\mathbf{y}_{_{S_1}}(t) \\
	\vdots \\
	\mathbf{y}_{_{S_N}}(t)
	\end{pmatrix}. \label{eq:state}
\end{align} 
$\mathbf{y}_{_{P}}(t)$ refers to the central planet's state. $\mathbf{y}_{_{M}}(t)$ is the moons' state vector with respect to the central planet, of size $6\times n$ with $n$ the number of moons. Finally, $\mathbf{y}_{_{S_i}}(t)$ represents the state of the $i^\mathrm{th}$ spacecraft with respect to its arc-wise central moon $m_{i,j}$. $N$ is the number of spacecraft involved in the estimation (equal to 2 in our analyses). 

The full initial state vector to be estimated can thus be written as follows:
\begin{align}
	\mathbf{y}_0 = \begin{pmatrix}
		\mathbf{y}_{_{P}}(t_0) \\
		\mathbf{y}_{_{M}}(t_0) \\
		\mathbf{y}_{_{S_1}}(\mathbf{t}_{_{S_1}}) \\
		\vdots \\
		\mathbf{y}_{_{S_N}}(\mathbf{t}_{_{S_N}})
	\end{pmatrix} \label{eq:initialState}
\end{align}
where $\mathbf{t}_{_{S_i}}$ contains the arc-wise reference epochs for spacecraft $S_i$. The spacecraft's states being estimated in an arc-wise manner, their initial state vector can be further expanded: 
\begin{align}
\mathbf{y}_{_{S_i}}(\mathbf{t}_{_{S_i}}) = \begin{pmatrix}
	\mathbf{y}_{_{S_i,1}}(t_{_{S_i,1}}) \\
	\vdots \\
	\mathbf{y}_{_{S_i,a_1}}(t_{_{S_i,a_i}})
\end{pmatrix},
\end{align}
with $a_i$ the number of arcs over which spacecraft $S_i$ is propagated, and $t_{_{S_i,j}}$ the reference epoch of arc $j$.

However, the equations of motion and variational equations are generally propagated in a single reference frame. States expressed in this global propagatation reference frame will be designed by the superscript $\star$ in the following. In contrast to Eq. \ref{eq:state}, the \textit{propagated} state can be defined as 
\begin{align}
	\mathbf{y}^\star(t) = 
	\begin{pmatrix}
	\mathbf{y}^\star_{_{P}}(t) \\
	\mathbf{y}^\star_{_{M}}(t) \\
	\mathbf{y}^\star_{_{S_1}}(t) \\
	\vdots \\
	\mathbf{y}^\star_{_{S_N}}(t)
	\end{pmatrix}.
\end{align}
The propagated and estimated states can be related using the following:
\begin{align}
	\mathbf{y}_{_{P}}(t) &= \mathbf{y}^\star_{_{P}}(t)\\
	\mathbf{y}_{_{M}}(t) & = \mathbf{y}^\star_{_{M}}(t) + \mathbf{y}^\star_{_{P}}(t) \\
	\mathbf{y}_{_{S_i}}(t) &= \mathbf{y}^\star_{_{S_i}}(t) + \mathbf{y}^\star_{_{m_{i,j}}}(t), \text{with } t\in\left[t_{_{S_i,j}} ; \tilde{t}_{_{S_i,j}}\right] \label{eq:stateConversion}
\end{align}
with $m_{i,j}$ the central moon of spacecraft $S_i$ during arc $j$. $t_{_{S_i,j}}$ and $\tilde{t}_{_{S_i,j}}$ respectively represent the start and end times of arc $j$ for spacecraft $S_i$.

In our analyses, the covariance matrix $\mathbf{P}$ describes the uncertainties and correlations of the state parameters with respect to their respective central bodies, according to Eq. \ref{eq:state}. To compute $\mathbf{P}$ using Eq. \ref{eq:cov}, the observation matrix $\mathbf{H}$ must first be computed to obtain the covariance matrix: 
\begin{align}
	\mathbf{H}(\mathbf{q}) = \frac{\partial\mathbf{h}(\mathbf{q})}{\partial\mathbf{q}},
\end{align}
with $\mathbf{h}$ the observations vector and $\mathbf{q}$ the parameters vector, which can be written as $\mathbf{q} = \left[\mathbf{y}_0,\mathbf{p}\right]^\mathrm{T}$. $\mathbf{y}_0$ is the initial state given by Eq. \ref{eq:initialState}, and $\mathbf{p}$ contains the non-state parameters. Focusing on the estimation of the initial state, we can then write, for a single observation:
\begin{align}
	\frac{\partial h(\mathbf{q})}{\partial \mathbf{y}_0} & =\frac{\partial h(\mathbf{q})}{\partial \mathbf{y}_0^\star}\frac{\partial \mathbf{y}_0^\star}{\partial \mathbf{y}_0},\\
	&= \frac{\partial h(\mathbf{q})}{\partial \mathbf{y}^\star(t)} \mathbf{\Phi}^\star(t,t_0,\mathbf{t}_{_{S_1}}, \text{...}, \mathbf{t}_{_{S_N}})\frac{\partial\mathbf{y}_0^\star}{\partial\mathbf{y}_0}.
\end{align}
$\frac{\partial h(\mathbf{q})}{\partial \mathbf{y}^\star(t)}$ and $\mathbf{\Phi}^\star(t, t_0,\mathbf{t}_{_{S_1}}, \text{...}, \mathbf{t}_{_{S_N}})$ can be computed after propagating the variational equations with respect to the global propagation reference frame, while $\frac{\partial\mathbf{y}_0^\star}{\partial\mathbf{y}_0}$ can be derived from Eq. \ref{eq:stateConversion}. It must be noted that $\mathbf{\Phi}^\star(t,t_0, \mathbf{t}_{_{S_1}}, \text{...}, \mathbf{t}_{_{S_N}})$ is equivalent to the state transition matrix $\mathbf{\Phi}(t,t_0)$ in Eq. \ref{eq:propCov}. A less detailed notation was indeed adopted in the core part of the paper for the sake of conciseness. 

Finally, the propagated covariance is given by
\begin{align}
	\mathbf{P}(t) = \left(\frac{\partial\mathbf{y}(t)}{\partial\mathbf{y}_0}\right)\mathbf{P}\left(\frac{\partial\mathbf{y}(t)}{\partial\mathbf{y}_0}\right)^\mathrm{T}. \label{eq:propCov1}
\end{align}
The partials in Eq. \ref{eq:propCov1} must again be re-written with respect to the \textit{propagated}, and not \textit{estimated}, state:
\begin{align}
	\frac{\partial\mathbf{y}(t)}{\partial\mathbf{y}_0} &= \frac{\partial\mathbf{y}(t)}{\partial\mathbf{y}^\star(t)} \frac{\partial \mathbf{y}^\star(t)}{\partial \mathbf{y}^\star_0} \frac{\partial \mathbf{y}^\star_0}{\partial \mathbf{y}_0}\\
	& = \frac{\partial\mathbf{y}(t)}{\partial\mathbf{y}^\star(t)} \mathbf{\Phi}^\star(t,t_0) \frac{\partial \mathbf{y}^\star_0}{\partial \mathbf{y}_0}.
	 \label{eq:propCov2}
\end{align}
Again, $\frac{\partial\mathbf{y}(t)}{\partial\mathbf{y}^\star(t)}$ and $\frac{\partial \mathbf{y}^\star_0}{\partial \mathbf{y}_0}$ can be extracted from Eq. \ref{eq:stateConversion}.

This small model extension completes the coupled estimation formulation provided in \cite{fayolle2022}. The main addition is the possibility to include the central planet's state in the estimation, which allows us to account for the Jovian ephemeris uncertainty in our analyses (Section \ref{sec:parameters}). The proposed implementation can however be applied to any planetary system and is versatile enough to accommodate any number of moons or spacecraft. 

\setcounter{figure}{0}
\setcounter{table}{0}

\section{Tidal dissipation estimates} \label{appendix:tidalDissipation}

This appendix presents the uncertainties in the inverse tidal quality factors for all Galilean satellites and Jupiter. $Q_{m}$ refers to the tidal quality factor of moon $m$, while $1/Q_{J,m}$ denotes that of Jupiter at the frequency of moon $m$. Table \ref{tab:baselineErrorsInvQ} reports the formal errors obtained with the baseline solution (no VLBI), and when adding PRIDE VLBI measurements (for both the poor and good VLBI noise budget scenarios). As underlined in Section \ref{sec:dynamicalModels}, it is essential to keep in mind that these estimates are solely based on the signatures of tidal dissipation in the moons' orbits, and do not account for tides sensed by the spacecraft. This choice was motivated by the focus of our paper on the Galilean moons' ephemerides and on the physical effects encoded in their dynamics. Dedicated studies investigating the effects of tides on both the moons and spacecraft's orbits can be found in e.g., \cite{cappuccio2020,demarchi2022,magnanini2023}.
As mentioned in Section \ref{sec:resultsVlbiGlobal}, no improvement is noticeable for the dissipation in either Io, Europa, or Ganymede, as well as for the dissipation in Jupiter at these moons' frequencies. A very limited uncertainty reduction can be observed for Callisto (and the corresponding estimation of Jupiter's estimation), which can be related to the small improvement of Callisto's along-track component attainable with VLBI (see Section \ref{sec:resultsVlbiGlobal}).

\begin{table*}[ht!]
	\caption{Formal uncertainties of the inverse of the tidal quality factors for each moon ($1/Q_{m}$), as well as for Jupiter at each moon's frequency ($1/Q_{J,m}$). The formal errors correspond to the baseline solution, without VLBI.}
	\label{tab:baselineErrorsInvQ}
    \small
	\centering
	\begin{tabular}{l l l c c c }
		& & \textbf{Io} & \textbf{Europa} & \textbf{Ganymede} & \textbf{Callisto} \\ \hline
		\multirow{3}{*}{$\sigma(1/Q_{J,m})$ [-]} & no VLBI & $3.5\times10^{-6}$ & $1.8\times10^{-3}$ & $2.5\times10^{-4}$ & $5.0\times10^{-2}$\\
        & poor VLBI & $3.5\times10^{-6}$ & $1.8\times10^{-3}$ & $2.5\times10^{-4}$ & $5.0\times10^{-2}$\\
        & good VLBI & $3.5\times10^{-6}$ & $1.8\times10^{-3}$ & $2.5\times10^{-4}$ & $4.8\times10^{-2}$ \\
        \hline
		\multirow{3}{*}{$\sigma(1/Q_m)$ [-]} & no VLBI & $7.7\times10^{-3}$& $1.1\times10^{-2}$& $1.7\times10^{-2}$& $4.0\times10^{-1}$\\
        & poor VLBI & $7.7\times10^{-3}$& $1.1\times10^{-2}$& $1.7\times10^{-2}$& $3.7\times10^{-1}$ \\
        & good VLBI & $7.7\times10^{-3}$& $1.1\times10^{-2}$& $1.6\times10^{-2}$& $3.3\times10^{-1}$ \\
		\hline
	\end{tabular}
\end{table*}

\setcounter{figure}{0}
\setcounter{table}{0}

\section{Multi-spacecraft VLBI contribution to the global solution with different estimation setups} \label{appendix:msVlbiDifferentSetups}

\begin{table*}[ht!]
	\caption{Improvement in \textbf{averaged} formal position uncertainties (percentage) with respect to the solution obtained with no VLBI, for various multi-spacecraft VLBI tracking scenarios. The position errors are computed in the RTN frame, and only improvements larger than 5\% are reported. We chose this low threshold value for this table to better support our discussion on the influence of the tracking configuration).}
	\label{tab:msVlbiDifferentSetups}
	\centering
	\resizebox{\textwidth}{!}{\begin{tabular}{ c  c | c | c c c | c c c | c c c | c c c | c c c  }
			\multicolumn{2}{c|}{\multirow{2}{*}{\textbf{Tracking arc}}} & \textbf{Noise} & \multicolumn{3}{c|}{\textbf{Jupiter}} & \multicolumn{3}{c|}{\textbf{Io}} & \multicolumn{3}{c|}{\textbf{Europa}} & \multicolumn{3}{c|}{\textbf{Ganymede}} & \multicolumn{3}{c}{\textbf{Callisto}} \\
			
			\multicolumn{2}{c|}{} & \textbf{budget} & R & T & N & R & T & N & R & T & N & R & T & N & R & T & N \\ 
			\hline
			
			\multicolumn{18}{c}{} \\
			\multicolumn{18}{l}{\textbf{Range and VLBI biases estimated}} \\ \hline
				
			$2\times$8h & arc bounds & poor & 9.9 & 7.5 & 44.1 & 8.6 & 7.8 & 8.4 & 8.9 & 7.6 & 22.5 & 6.6 & 7.1 & 9.4 & 9.4 & 9.3 & 9.8 \\ 
			16h & mid-arc & poor & 10.7 & 7.2 & 42.1 & 12.8 & 10.6 & 13.5 & 9.3 & 10.1 & 26.0 & 7.4 & 8.3 & 11.1 & 9.4 & 9.4 & 9.5 \\ \hline
			
			\multicolumn{2}{c|}{Full tracking} & poor & 18.7 & 11.9 & 51.2 & 16.1 & 16.6 & 12.2 & 16.3 & 13.6 & 47.4 & 15.0 & 15.4 & 28.4 & 15.7 & 18.0 & 21.5 \\ \hline
					
			$2\times$8h & arc bounds & good & 16.1 & 13.4 & 58.0 & 12.4 & 12.3 & 10.9 & 12.3 & 10.5 & 35.9 & 10.1 & 10.9 & 14.2 & 13.5 & 15.1 & 18.8 \\
			16h & mid-arc & good & 18.8 & 13.6 & 58.9 & 15.2 & 14.5 & 15.4 & 12.9 & 12.7 & 40.3 & 12.7 & 13.2 & 19.5 & 15.2 & 16.1 & 18.4 \\ \hline 
			
			\multicolumn{2}{c|}{Full tracking} & good & 28.5 & 20.1 & 67.6 & 19.2 & 21.1 & 15.2 & 21.3 & 16.4 & 59.1 & 24.5 & 22.4 & 41.5 & 21.8 & 27.4 & 37.5 \\ \hline			
			\multicolumn{18}{c}{} \\
			\multicolumn{18}{l}{\textbf{Range and VLBI biases considered}} \\ \hline
			
			$2\times$8h & arc bounds & poor & 16.9 & 16.1 & 55.8 & 8.0 & 7.0 & 17.0 & - & 11.2 & 34.2 & 10.2 & 13.5 & 11.4 & 23.0 & 23.4 & 23.0 \\ 
			16h & mid-arc & poor & 22.3 & 19.5 & 59.5 & - & 7.6 & 19.7 & - & 11.2 & 36.1 & 14.0 & 15.8 & 13.7 & 25.6 & 26.2 & 25.1 \\ \hline
			
			\multicolumn{2}{c|}{Full tracking} & poor & 38.5 & 31.8 & 71.7 & 12.3 & 13.9 & 20.5 & 12.3 & 17.5 & 54.7 & 27.7 & 26.8 & 35.1 & 37.1 & 40.4 & 43.4 \\ \hline
					
			$2\times$8h & arc bounds & good & 24.4 & 23.9 & 63.2 & 10.9 & 11.2 & 21.5 & - & 14.8 & 46.4 & 15.7 & 18.4 & 19.2 & 28.6 & 29.4 & 30.2 \\
			16h & mid-arc & good & 30.4 & 27.2 & 67.3 & 7.0 & 11.8 & 23.0 & - & 14.6 & 49.5 & 20.6 & 21.2 & 21.9 & 31.3 & 32.4 & 32.9 \\ \hline
			
			\multicolumn{2}{c|}{Full tracking} & good & 47.8 & 38.5 & 79.6 & 17.6 & 19.1 & 26.6 & 20.0 & 21.7 & 68.6 & 34.5 & 32.4 & 51.2 & 42.5 & 47.4 & 52.8 \\	\hline
	\end{tabular}}
\end{table*}

As discussed in Section \ref{sec:resultsMsVlbiGlobal}, our choice of baseline estimation setup - estimating both range and VLBI biases - leads to a conservative estimate of the global solution improvement attainable with multi-spacecraft VLBI. For the sake of completeness, we ran the same analysis while including range and VLBI biases as consider parameters. The results are reported in Table \ref{tab:msVlbiDifferentSetups} for a limited number of tracking configurations, and indeed show larger improvements than with the baseline setup. Only Io  and Europa's in-plane position uncertainties slightly degrade when adding range and VLBI biases as consider parameters instead of estimating them. This can be explained by the fact that Europa's solution, and thus indirectly Io's, strongly rely on Europa Clipper radio-science (i.e. Doppler only, Section \ref{sec:simulatedTracking}). The baseline solution for these components is thus less sensitive to range biases, and only the VLBI contribution is notably affected by the change of estimation setup. However, for the rest of the state parameters, the VLBI improvement strengthens when the observation biases are not estimated, due to the deterioration of the baseline solution. Overall, as mentioned in Section \ref{sec:resultsMsVlbiGlobal}, our results indicate that a stronger contribution could possibly be expected from multi-spacecraft VLBI measurements, depending on the accuracy of the baseline solution.

\setcounter{figure}{0}
\setcounter{table}{0}

\section{Multi-spacecraft VLBI contribution to the global solution for different sets of flyby combinations} \label{appendix:msVlbiFlybysCombinations}

\begin{table*}[ht!]
	\caption{Improvement in \textbf{averaged} formal position uncertainties (percentage) with respect to the baseline solution. The results are obtained in the arc bounds tracking configuration (tracking arcs of 8h), exploiting different subsets of the flyby combinations in Table \ref{tab:msVlbiOptions}. Only improvements larger than 5\% are reported.}
	\label{tab:msVlbiFlybysCombinations}
	\centering
	\resizebox{\textwidth}{!}{\begin{tabular}{ c  | c | c c c | c c c | c c c | c c c | c c c  }
			
			\textbf{Time limit} & \textbf{Noise} & \multicolumn{3}{c|}{\textbf{Jupiter}} & \multicolumn{3}{c|}{\textbf{Io}} & \multicolumn{3}{c|}{\textbf{Europa}} & \multicolumn{3}{c|}{\textbf{Ganymede}} & \multicolumn{3}{c}{\textbf{Callisto}} \\
			\textbf{in-between flybys}& \textbf{budget} & R & T & N & R & T & N & R & T & N & R & T & N & R & T & N \\ 
			\hline
			
			\multicolumn{1}{c|}{3 days (Table \ref{tab:msVlbiGlobal})} & poor & 9.9 & 7.5 & 44.1 & 	8.6 & 7.8 & 8.4 & 8.9 & 7.6 & 22.5 & 6.6 & 7.1 & 9.4 & 9.4 & 9.3 & 9.8 \\ 
			
			1 day & poor & 5.9 & - & 35.4 & - & - & 7.0 & - & 4.8 & 13.4 & - & - & - & 5.4 & - & - \\
			\hline		
			
			\multicolumn{17}{c}{} \\ \hline
			
			\multicolumn{1}{c|}{3 days (Table \ref{tab:msVlbiGlobal})} & good & 16.1 & 13.4 & 58.0 & 12.4 & 12.3 & 10.9 & 12.3 & 10.5 & 35.9 & 10.1 & 10.9 & 14.2 & 13.5 & 15.1 & 18.8 \\
			
			1 day & good & 11.3 & 8.0 & 52.9 & 5.0 & 7.0 & 8.8 & 7.1 & 7.1 & 26.8 & 6.2 & 7.3 & 9.2 & 8.4 & 5.9 & - \\	
			\hline	
	\end{tabular}}
\end{table*}

11 flyby combinations were identified as representing promising opportunities to perform multi-spacecraft VLBI tracking (Section \ref{sec:msVlbi}, Table \ref{tab:msVlbiOptions}). An upper threshold of three days between each JUICE flyby and the closest Europa Clipper flyby was applied. However, as discussed in Section \ref{sec:resultsMsVlbiGlobal}, such an elapsed time in-between the two flybys would require extending the nominal tracking sessions and thus be more resource-demanding. Interestingly, the JUICE and Europa Clipper flybys are planned less than one day apart for three combinations (Table \ref{tab:msVlbiOptions}), such that multi-spacecraft VLBI could be acquired at minimal expense, without extending the nominal tracking arcs. Table \ref{tab:msVlbiFlybysCombinations} compares the solution improvement achieved when simulating multi-spacecraft VLBI either during all 11 combinations, or just during the above-mentioned three combinations with close flybys. The results are discussed in Section \ref{sec:resultsMsVlbiGlobal}.

\end{document}